\documentclass[twocolumn]{aastex631}

\usepackage{amsmath}
\usepackage{mathtools}
\usepackage{tikz}
\usetikzlibrary{arrows,shapes,backgrounds,calc,positioning,tikzmark,patterns,automata,decorations.markings}
\tikzset{fontscale/.style={font=\relsize{#1}}}
\tikzset{->-/.style={decoration={
  markings,
  mark=at position #1 with {\arrow{>}}},postaction={decorate}}}
\tikzset{-<-/.style={decoration={
  markings,
  mark=at position #1 with {\arrow{<}}},postaction={decorate}}}

\tikzset{cross/.style={cross out,draw,minimum size=2*(#1-\pgflinewidth),inner sep=0pt, outer sep=0pt}}

\tikzset{
  pics/carc/.style args={#1:#2:#3}{
    code={
      \draw[pic actions] (0,0) -- (#1:#3) arc(#1:#2:#3) -- cycle;
    }
  }
}
\usepackage{stackrel}

\newcommand{\curl}{\operatorname{curl}}
\newcommand{\efnum}[2]{{#1}\cdot 10^{#2}}

\definecolor{refcorrcol}{RGB}{100,0,0}
\newcounter{inrefcorr}
\newcommand{\refcorr}[1]{\stepcounter{inrefcorr}\ifmmode{\color{refcorrcol}#1}\else\textbf{#1}\fi\addtocounter{inrefcorr}{-1}}
\everymath{\ifnum\value{inrefcorr}>0\color{refcorrcol}\fi}

\graphicspath{{./}{figures/}}

\submitjournal{ApJ}

\shorttitle{Spheromak tilting}
\shortauthors{Asvestari et al.}

\begin{document}

\title{The spheromak tilting and how it affects modelling coronal mass ejections}

\correspondingauthor{Eleanna Asvestari}
\email{eleanna.asvestari@helsinki.fi}

\author[0000-0002-6998-7224]{Eleanna Asvestari}
\affiliation{Faculty of Science, Department of Physics, University of Helsinki, Gustaf H{\"a}llstr{\"o}min katu 2,
Helsinki, Finland}

\author[0000-0003-2617-4319]{Tobias Rindlisbacher}
\affiliation{Faculty of Science, Department of Physics, University of Helsinki, Gustaf H{\"a}llstr{\"o}min katu 2,
Helsinki, Finland}
\affiliation{Albert Einstein Center for Fundamental Physics, Institute for Theoretical Physics, University of Bern,  \\
Sidlerstrasse 5, CH-3012 Bern, Switzerland}

\author[0000-0003-1175-7124]{Jens Pomoell}
\affiliation{Faculty of Science, Department of Physics, University of Helsinki, Gustaf H{\"a}llstr{\"o}min katu 2,
Helsinki, Finland}

\author[0000-0002-4489-8073]{Emilia K. J. Kilpua}
\affiliation{Faculty of Science, Department of Physics, University of Helsinki, Gustaf H{\"a}llstr{\"o}min katu 2,
Helsinki, Finland}

\begin{abstract}

Spheromak type flux ropes are increasingly used for modelling coronal mass ejections (CMEs). Many models aim in accurately reconstructing the magnetic field topology of CMEs, considering its importance in assessing their impact on modern technology and human activities in space and on ground. However, so far there is little discussion about how the details of the magnetic structure of a spheromak affect its evolution through the ambient field in the modelling domain, and what impact this has on the accuracy of magnetic field topology predictions. If the spheromak has its axis of symmetry (geometric axis) at an angle with respect to the direction of the ambient field, then the spheromak starts rotating so that its symmetry axis finally aligns with the ambient field. When using the spheromak in space weather forecasting models this tilting can happen already during insertion and significantly affects the results. In this paper we highlight this issue previously not examined in the field of space weather and we estimate the angle by which the spheromak rotates under different conditions. To do this we generated simple purely radial ambient magnetic field topologies (weak/strong positive/negative) and inserted spheromaks with varying initial speed and tilt, and magnetic helicity sign. We employ different physical and geometric criteria to locate the magnetic centre of mass and axis of symmetry of the spheromak. We confirm that spheromaks rotate in all investigated conditions and their direction and angle of rotation depend on the spheromak's initial properties and ambient magnetic field strength and orientation.

\end{abstract}


\section{Introduction}\label{sec:intro}

The ability to model coronal mass ejections (CMEs) as magnetic structures is of great importance within the space weather community. Depending on their magnetic structure, certain Earth-directed CME events can affect the near-Earth environment much stronger than others, and thus have a greater impact on human activity in space and on ground. These CMEs are called geoeffective and their key characteristic is a strong southward $B_z$-component, which is anti-parallel to Earth's magnetic field \citep{kilpua_geoeffective_2017}. The result of this anti-parallel configuration of the CME-Earth magnetic field components is day-side reconnection resulting in an effective transfer of solar wind energy, mass, and momentum into the Earth's magnetosphere. During such periods significant disturbances occur in the geomagnetic field and in the radiation environment of Earth that can, for example damage spacecraft orbiting Earth, affect navigation systems, and induce currents in power grids on ground.

While observations have led to the general consensus that CMEs consist of twisted flux ropes that have foot-points which remain attached to the Sun, their global magnetic morphology is not yet fully resolved. For CME modelling purposes, magnetohydrodynamic (MHD) simulation models often employ spheromak-type magnetic field configurations. The spheromak is a force-free, axisymmetric MHD equilibrium configuration where a twisted magnetic field fills a spherical volume and confines the plasma. An introduction to spheromaks and different stationary solutions relevant to lab-implementations can be found in \citet{bellan_fundamentals_2000}. Spheromak-type CME implementations in MHD simulation models can be divided in two broad categories, the ones in which the spheromak remains anchored to the inner boundary located in the low corona \citep[see for example,][]{gibson_time-dependent_1998, manchester_three-dimensional_2004, manchester_modeling_2004, manchester_flux_2014, manchester_simulation_2014, lugaz_numerical_2005, singh_data-constrained_2018, singh_application_2020, singh_modified_2020, jin_data-constrained_2017} and those in which it is fully inserted into the middle or upper corona and therefore retains its magnetically confined spherical nature \citep[see for example,][]{kataoka_three-dimensional_2009, shiota_magnetohydrodynamic_2016, verbeke_evolution_2019, scolini_cmecme_2020, asvestari_multispacecraft_2021}. 
Note that both of these spheromak types used in the modelling of magnetised CMEs are different from typical lab-implementations of spheromaks: while the latter are stationary, the spheromaks used in CME models are expanding, non-stationary structures. However, both, lab- and CME-spheromaks have in common that they carry a magnetic moment, which, when exposed to an ambient magnetic field, aims to align itself with this ambient field. A magnetically isolated CME-spheromak could therefore during its evolution undergo a change in orientation of its entire magnetic field structure. The possibility of such a rotational component in the dynamic evolution of model-CMEs has been addressed only briefly by the aforementioned studies, but could be significant for explaining agreement or disagreement between modelled and observed \textit{in situ} magnetic field topology of CMEs at larger heliodistances~\citep[]{kataoka_three-dimensional_2009}.

A related phenomenon is the tilting instability of a spheromak in an ambient magnetic field, which can occur if the magnetic moment of the spheromak is anti-aligned with the ambient field. The instability, when triggered, then results in a rotation --tilting-- of the spheromak, aiming to lower its potential magnetic energy \citep[][]{rosenbluth_bussac_1979, bellan_fundamentals_2000, mehta_spheromak_tilting_2021}. \citet{shiota_mhd_tilting_2010} modelled an active region eruption by embedding a spheromak in a global dipole field. The authors reported that the spheromak experienced rotation due to the presence of a torque due to magnetic force, similar to the spheromak tilting instability reported in experiments of laboratory plasmas \citep{sato_hayashi_1983}. Such model results are in accordance to observations showing that some CMEs rotate in the low solar corona \citep{yurchyshyn_obs_rotate_2008}.

Although CMEs can undergo significant kinking, rotation, and deflection upon eruption and during the early evolution close to the Sun \citep[e.g.,][]{kay_global_2015, kay_heliocentric_2015, heinemann_cmehss_2019}, these are less frequently reported evolutionary aspects of CMEs further out in interplanetary space \citep[][]{Isavnin_evolution_2014}.  White light observations between 1.5-30~$R_{Sun}$ suggest that most CMEs seize to undergo significant rotation or deflection at larger heliocentric distances and instead seem to expand in a self-similar manner as they propagate further away from the Sun \citep[e.g.,][and references within]{colaninno_analysis_2006, demoulin_causes_2009, balmaceda_expansion_2020}. As we show in this study, this \textit{almost rotation-free} expansion in interplanetary space might often not be accurately reproduced by spheromak CME models. 

In this paper we demonstrate, using the EUHFORIA MHD heliospheric model \citep{pomoell_euhforia_2018}, that a spheromak injected into the inner-heliospheric solar wind experiences tilting and deflection as it evolves to larger heliocentric distances, which results in a gradual change of orientation of the spheromak's axis of symmetry and a slight alteration of its trajectory. We locate the magnetic centre of mass of the modelled spheromaks (centre of mass with respect to the magnetic field energy density) and their axis of symmetry, and investigate the response of spheromaks with different initial states to the ambient magnetic field. To study this we generated simple ambient magnetic field topologies, as described in Section \ref{sec:methods}.
More complex topologies such as ones including heliospheric current sheet crossings and high speed stream structures, are not addressed in this paper; however, we intend to investigate them in future work.

Section \ref{sec:spheromakandtilting} provides a more in-depth discussion of the physics behind the spheromak tilting instability. In section \ref{sec:methods} we describe the different simulation setup considered in this analysis, while in section \ref{sec:results} we provide in more detail our analysis and main findings. Details on the process of locating the spheromak in the three--dimensional modelling domain, its magnetic centre of mass, and its symmetry axis are given in Appendix \ref{apsec:spheromak_monitoring}. Appendix \ref{apsec:coordinate_transform} explains the coordinate transformation used for visualizing the located spheromaks, while Appendix \ref{apsec:torque_and_force_accuracy} describes in more detail the torque and net force the LFF spheromak experiences when subject to a background magnetic field. A summary of our key results, a brief discussion on the importance of these findings for the space weather community, and our future steps in investigating the spheromak insertion and evolution are presented in section \ref{sec:conclusion}.

\section{Spheromak and tilting instability} \label{sec:spheromakandtilting}

\begin{figure}[htb]
\centering
\includegraphics[height=0.66\linewidth,keepaspectratio]{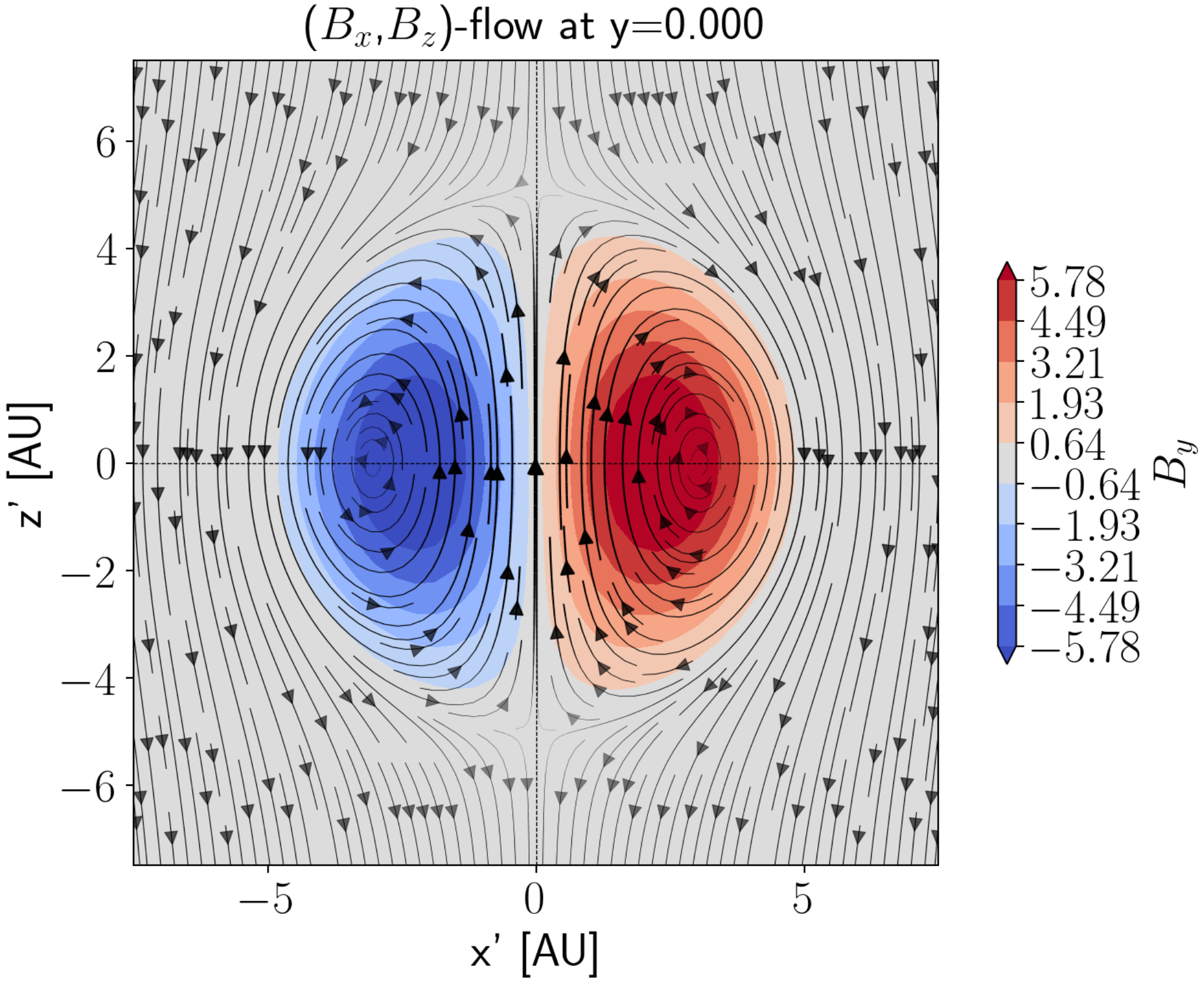}\\[10pt]
\includegraphics[height=0.66\linewidth,keepaspectratio]{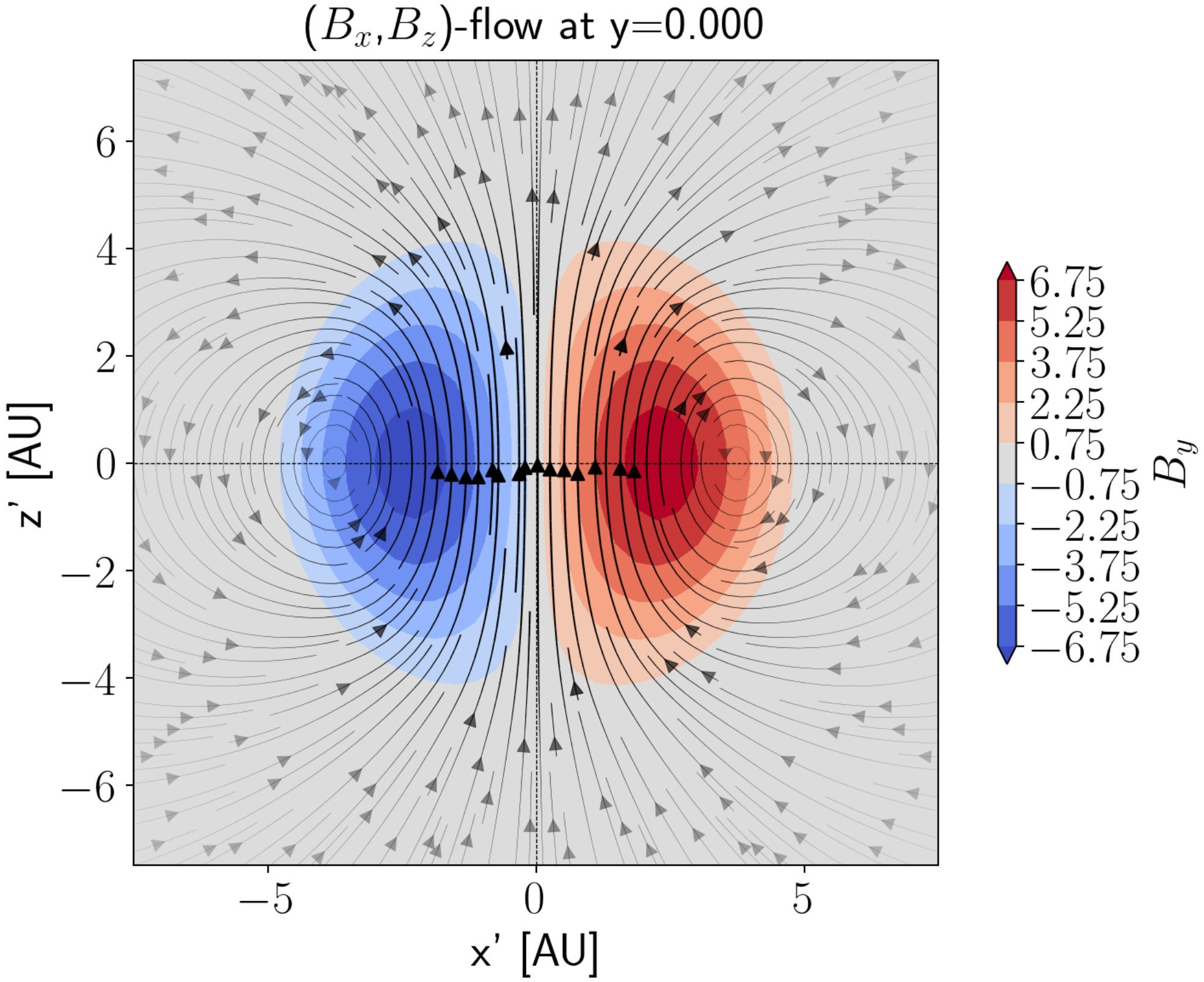}
\caption{Upper panel: total magnetic field of a spherically shaped ideal spheromak in a constant magnetic background field that is pointing in negative $\hat{z}$-direction.\\ 
Lower Panel: magnetic field of the same spheromak as above, but with the constant background field subtracted, namely it shows the magnetic field that is merely produced by the current distribution inside the spheromak. This resembles the magnetic field of a current loop dipole.}
\label{fig:idealspheromak}
\end{figure}
In this section we review two types of static, linear force free (LFF) spheromak solutions and how they are related to the tilting instability \citep[][chapter 10.2-10.3]{bellan_fundamentals_2000} and to the application of LFF spheromaks as CME models for MHD simulations of the inner heliosphere.

What we call LFF spheromak here, is an axisymmetric solution of the force-free equation,
\begin{equation}
\curl(\mathbf{B})=\lambda \mathbf{B}\ \label{eq:forcefreeeq}
\end{equation}
that (approximately) describes the force balance in a stationary $\beta\approx 0$ (magnetically dominated) ideal MHD-plasma, with $\lambda$ having a constant, non-zero value only within a spherically shaped region, $V$, which we identify to be the volume of the spheromak. Outside of $V$, $\lambda$ vanishes.\\

\subsection{LFF spheromak in homogeneous background field}\label{ssec:lffsph_homogeneous_background}
The magnetic field $\mathbf{B}$ in equation \eqref{eq:forcefreeeq} can be thought of as consisting of two parts:
\begin{equation}
\mathbf{B}=\mathbf{B}_{\mathrm{bs}}+\mathbf{B}_{\mathrm{vac}}\ 
\end{equation}
where $\mathbf{B}_{\mathrm{bs}}$ represents the magnetic field produced by the current distribution $\mathbf{J}=\curl(\mathbf{B})/\mu_0$ inside $V$ (Biot-Savart law):
\begin{equation}
\mathbf{B}_{\mathrm{bs}}(\mathbf{r})=\frac{\mu_0}{4\,\pi}\int\frac{\mathbf{J}(\mathbf{r}')\times (\mathbf{r}-\mathbf{r}')}{|\mathbf{r}-\mathbf{r}'|^3}\mathrm{d}V'\ \label{eq:bbiotsavart}
\end{equation}
and $\mathbf{B}_{\mathrm{vac}}$ represents the background vacuum field with $\curl(\mathbf{B}_{\mathrm{vac}})=0$. The splitting of the magnetic field into a background contribution and a contribution that can be considered to be part of the spheromak itself, is necessary to be able to identify the magnetic net forces that act on the spheromak as a whole. The part of the magnetic field that is due to the current density of the spheromak itself can (to some extent) be considered rigidly coupled to the spheromak and cannot exert any magnetic net force or net torque on itself. If such a magnetic net force or a net torque act on the spheromak, it is the background part of the magnetic field that is responsible.

An example of a stationary spheromak solution in a homogeneous background field $\mathbf{B}_{\mathrm{vac}}=-B_0\,\mathbf{\hat{z}}$, pointing in negative z-direction, is shown in Figure~\ref{fig:idealspheromak}. The upper panel shows $\mathbf{B}=\mathbf{B}_{\mathrm{bs}}+\mathbf{B}_{\mathrm{vac}}$, and the lower panel shows the contribution $\mathbf{B}_{\mathrm{bs}}$, which qualitatively looks like the field of a current-loop magnetic dipole.
The spheromak solution used in this figure has been known for many years~\citep{rosenbluth_bussac_1979} and was recently reconsidered in \citet{mehta_spheromak_tilting_2021}. In terms of spherical coordinates $(r,\theta,\phi)=(\text{radius},\text{colatitude},\text{longitude})$  with origin in the centre of $V$ and using normalized coordinate tangent vectors $(\hat{r},\hat{\theta},\hat{\phi})$ as vector basis, the $\mathbf{B}$-field is in the interior of $V$ described by:
\begin{subequations}\label{eq:lffinternalsolution}
\begin{align}
B^{\textrm{in}}_r&=\frac{2 B_0^{\mathrm{in}}}{\lambda\,r}\,j_{1}(\lambda\,r) \cos(\theta)\\
B^{\textrm{in}}_{\theta}&=-\frac{B_0^{\mathrm{in}}}{\lambda\,r} \frac{\partial (r\,j_1(\lambda\,r))}{\partial r} \sin(\theta)\\
B^{\textrm{in}}_{\phi}&=B_0^{\mathrm{in}}\,j_1(\lambda\,r) \sin(\theta)
\end{align}
\end{subequations}
while the exterior field is:
\begin{subequations}\label{eq:lffexternalsolution}
\begin{align}
B^{\textrm{ex}}_r&=B_0^{\mathrm{ex}} \left(1-\frac{r_0^3}{r^3}\right) \cos(\theta)\\
B^{\textrm{ex}}_{\theta}&=-B_0^{\mathrm{ex}} \left(1+\frac{r_0^3}{2\,r^3}\right) \sin(\theta)\\
B^{\textrm{ex}}_{\phi}&=0
\end{align}
\end{subequations}
with $j_{n}(s)$ being the spherical Bessel function of the first kind of order $n$, $r_0$ the radius of the spherical region $V$, and $\lambda$ is determined by requiring $B^{\mathrm{in}}_r$ to vanish on the boundary of $V$. The latter condition yields infinitely many solutions, so-called \textit{Taylor states}, as $j_{1}(s)$ has infinitely many zeros as function of $s$. We restricted ourselves here to the lowest Taylor state, corresponding to the first non-trivial zero of $j_{1}(s)$, given by $s=s_1\approx 4.49341$. While having $B^{\mathrm{in}}_r=0$ on the boundary of $V$ requires also $B^{\mathrm{in}}_{\phi}$ to be zero there, $B^{\mathrm{in}}_{\theta}$ will have a non-zero value
\begin{equation}
B^{\mathrm{in}}_{\theta}=-B_0^{\mathrm{in}}\,j_0(s_1)\, \sin(\theta)\ .
\end{equation}
In order to avoid the formation of surface currents, $\mathbf{B}$ must be continuous across the boundary of $V$, and $B^{\mathrm{ex}}_{\theta}$ in Equations~\eqref{eq:lffexternalsolution} should therefore match $B^{\mathrm{in}}_{\theta}$ from Equations~\eqref{eq:lffinternalsolution} for $r=r_0$. This determines the $B_0^{\mathrm{in}}$ in \eqref{eq:lffinternalsolution} in terms of $B_0^{\mathrm{ex}}=-B_0$ as:
\begin{equation}
B_0^{\mathrm{in}}=\frac{3}{2\,j_0(s_1)}  B_0^{\mathrm{ex}} .\label{eq:lffinternexternrel}
\end{equation}
This latter Equation~\eqref{eq:lffinternexternrel} requires $\mathbf{B}_{\mathrm{vac}}$ to be non-zero and anti-parallel to the magnetic moment,
\begin{equation}
\mathbf{M}=\frac{1}{2}\int(\mathbf{r}\times \mathbf{J}(\mathbf{r}))\,\mathrm{d}V\,=\,2\,\pi\,B_0\,r_0^3\,\mathbf{\hat{z}}\ \label{eq:sphmmagmoment}
\end{equation}
of the spheromak's current distribution $\mathbf{J}=\curl(\mathbf{B})/\mu_0$, unless the spheromak is trivial (meaning $\lambda=0$).
Note also that $\mathbf{B}^{\mathrm{in}}$ and $\mathbf{B}^{\mathrm{ex}}$ are both superpositions of the restrictions of $\mathbf{B}_{\mathrm{bs}}$ and $\mathbf{B}_{\mathrm{vac}}$ to either the interior or exterior of $V$. The background field is in both regions the same:
\begin{subequations}\label{eq:lffbgfield}
\begin{align}
B^{\textrm{in}}_{\mathrm{vac},r}=B^{\textrm{ex}}_{\mathrm{vac},r}&=-B_0 \cos(\theta)\\
B^{\textrm{in}}_{\mathrm{vac},\theta}=B^{\textrm{ex}}_{\mathrm{vac},\theta}&=B_0 \sin(\theta)\\
B^{\textrm{in}}_{\mathrm{vac},\phi}=B^{\textrm{ex}}_{\mathrm{vac},\phi}&=0
\end{align}
\end{subequations}
while $\mathbf{B}_{\mathrm{bs}}$ is in the interior of $V$ given by:
\begin{subequations}\label{eq:lffbsinternalsolution}
\begin{align}
B^{\textrm{in}}_{\mathrm{bs},r}&=-B_0\left(\frac{3}{j_0(s_1)}\frac{j_{1}(\lambda\,r)}{\lambda\,r}-1\right) \cos(\theta)\\
B^{\textrm{in}}_{\mathrm{bs},\theta}&=B_0\left(\frac{3}{2\,j_0(s_1)\,\lambda\,r} \frac{\partial (r\,j_1(\lambda\,r))}{\partial r}-1\right) \sin(\theta)\\
B^{\textrm{in}}_{\mathrm{bs},\phi}&=-B_0\,\frac{3}{2\,j_0(s_1)}\,j_1(\lambda\,r) \sin(\theta)
\end{align}
\end{subequations}
and in the exterior by:
\begin{subequations}\label{eq:lffbsexternalsolution}
\begin{align}
B^{\textrm{ex}}_{\mathrm{bs},r}&=B_0 \frac{r_0^3}{r^3} \cos(\theta)\\
B^{\textrm{ex}}_{\mathrm{bs},\theta}&=B_0 \frac{r_0^3}{2\,r^3} \sin(\theta)\\
B^{\textrm{ex}}_{\mathrm{bs},\phi}&=0\ ,
\end{align}
\end{subequations}
so that in the interior of $V$ we have:
\begin{equation}
\mathbf{B}^{\mathrm{in}}=\mathbf{B}^{\mathrm{in}}_{\mathrm{bs}}+\mathbf{B}^{\mathrm{in}}_{\mathrm{vac}}
\end{equation}
and similarly in the exterior of $V$:
\begin{equation}
\mathbf{B}^{\mathrm{ex}}=\mathbf{B}^{\mathrm{ex}}_{\mathrm{bs}}+\mathbf{B}^{\mathrm{ex}}_{\mathrm{vac}}\ .
\end{equation}

It is important to stress that this spheromak solution is only force free with respect to the total magnetic field $\mathbf{B}$. The field $\mathbf{B}_{\mathrm{bs}}$ on its own does not satisfy the force-free condition in Equation~\eqref{eq:forcefreeeq}. The background field $\mathbf{B}_{\mathrm{vac}}$ is responsible for balancing the Lorentz force ($\mathbf{F}_{\mathrm{L}}$) that $\mathbf{B}_{\mathrm{bs}}$ exerts on its own source current $\mathbf{J}$:
\begin{equation}
0=\mathbf{F}_{\mathrm{L}}=\mathbf{J}\times \mathbf{B}=\mathbf{J}\times \mathbf{B}_{\mathrm{bs}}+\mathbf{J}\times \mathbf{B}_{\mathrm{vac}}\ .
\end{equation}
Without the Lorentz force contribution from the background field, the contribution due to $\mathbf{B}_{\mathrm{bs}}$, namely $\mathbf{F}_{\mathrm{L},\mathrm{bs}}=\mathbf{J}\times \mathbf{B}_{\mathrm{bs}}$ would push the current density $\mathbf{J}$ of the spheromak outwards. 

\subsection{LFF spheromak supported by surface currents}\label{ssec:lffsph_surface_currents}
A non-trivial, stationary LFF spheromak solution of the form of Equations~\eqref{eq:lffinternalsolution} inside a volume $V$ can also be realised without a non-zero background vacuum field, provided the boundary of $V$ is highly conducting. The total exterior field, $\mathbf{B}^{\mathrm{ex}}$, can then be zero. As discussed in~\citet[][chapter 10.2]{bellan_fundamentals_2000}, the resulting discontinuity in the $B_{\theta}$ component of the magnetic field when crossing the boundary of $V$ will then induce a toroidal surface current on the boundary, which, if the interior spheromak solution is still as in Equation~\eqref{eq:lffinternalsolution}, is given by:
\begin{subequations}\label{eq:sfcurrent}
\begin{align}
I^{\mathrm{sf}}_{r} &= 0\\
I^{\mathrm{sf}}_{\theta} &=0\\
I^{\mathrm{sf}}_{\phi} &= \frac{3\,B_0\,\sin(\theta)}{2\,\mu_0}
\end{align}
\end{subequations}
and runs in opposite direction to the toroidal component of the current density $\mathbf{J}=\curl(\mathbf{B})/\mu_0$ in the interior of $V$. The surface current, Equation~\eqref{eq:sfcurrent}, produces in the interior of $V$ a with respect to $\mathbf{B}_{\mathrm{bs}}$ anti-aligned magnetic field, $\mathbf{B}_{\mathrm{sf}}$, analogous to $\mathbf{B}_{\mathrm{vac}}$ from Section~\ref{ssec:lffsph_homogeneous_background}.
The total $\mathbf{B}$-field that satisfies the force-free equation~\eqref{eq:forcefreeeq} is again a superposition 
\begin{equation}
\mathbf{B}=\mathbf{B}_{\mathrm{bs}}+\mathbf{B}_{\mathrm{sf}}\ \label{eq:btotsf}
\end{equation}
where $\mathbf{B}_{\mathrm{bs}}$ is as before the magnetic field produced by the current density $\mathbf{J}=\curl(\mathbf{B})/\mu_0$, which is non-zero only in the interior of $V$, and $\mathbf{B}_{\mathrm{sf}}$ is the field produced by the surface current, Equation~\eqref{eq:sfcurrent}, on the boundary of $V$. The field $\mathbf{B}_{\mathrm{sf}}$ can as $\mathbf{B}_{\mathrm{bs}}$ be computed using the Biot-Savart law:
\begin{equation}
\mathbf{B}_{\mathrm{sf}}(\mathbf{r})=\frac{\mu_0}{4\,\pi}\int_{\partial V}\frac{\mathbf{I}^{\mathrm{sf}}(\mathbf{r'})\times (\mathbf{r}-\mathbf{r'})}{\left|\mathbf{r}-\mathbf{r'}\right|^3}\mathrm{d}S' ,\label{eq:sfbsint}
\end{equation}
where $\partial V$ is the boundary of $V$ an $\mathrm{d}S'$ the surface element at position $\mathbf{r'}$ on $\partial V$. The result of the integration in Equation~\eqref{eq:sfbsint} is in the interior of $V$ given by
\begin{subequations}\label{eq:sfinternalsolution}
\begin{align}
B^{\textrm{in}}_{\mathrm{sf},r}&=-B_0 \cos(\theta)\\
B^{\textrm{in}}_{\mathrm{sf},\theta}&=B_0 \sin(\theta)\\
B^{\textrm{in}}_{\mathrm{sf},\phi}&=0\ 
\end{align}
\end{subequations}
and in the exterior of $V$ by
\begin{subequations}\label{eq:sfexternalsolution}
\begin{align}
B^{\textrm{ex}}_{\mathrm{sf},r}&=-B_0\frac{r_0^3}{r^3} \cos(\theta)\\
B^{\textrm{ex}}_{\mathrm{sf},\theta}&=-B_0\frac{r_0^3}{2\,r^3} \sin(\theta)\\
B^{\textrm{ex}}_{\mathrm{sf},\phi}&=0\ .
\end{align}
\end{subequations}
In terms of Cartesian coordinates, the interior solution from Equation~\eqref{eq:sfinternalsolution} is simply $\mathbf{B}_{\mathrm{sf}}=-B_0\,\mathbf{\hat{z}}$ and therefore indeed equal to the background field from Section~\ref{ssec:lffsph_homogeneous_background}. This has to be the case, in order to have the total B-field from Equation~\eqref{eq:btotsf}, which satisfies the force-free condition Equation~\eqref{eq:forcefreeeq}, being still given by Equations~\eqref{eq:lffinternalsolution}. In the exterior of $V$, $\mathbf{B}^{\mathrm{ex}}_{\mathrm{sf}}$ exactly cancels $\mathbf{B}^{\mathrm{ex}}_{\mathrm{bs}}$, so that the total field vanishes there and the force free equation is satisfied trivially.

It is interesting to note, that if this solution is indeed kept static, then the magnetic moment associated to the surface currents exactly cancels the magnetic moment in Equation~\eqref{eq:sphmmagmoment}.

\subsection{Tilting instability}\label{ssec:lffsph_torque_force_tilt}
The tilting instability can arise in the scenario discussed in Section~\ref{ssec:lffsph_homogeneous_background}, where the spheromak is located inside a background vacuum field, $\mathbf{B}_{\mathrm{vac}}$, with its magnetic moment $\mathbf{M}$ oriented anti-parallel to $\mathbf{B}_{\mathrm{vac}}$. The instability is caused by the fact that the torque exerted on $\mathbf{M}$ by $\mathbf{B}_{\mathrm{vac}}$ is given by (see Appendix~\ref{apsec:torque_and_force_accuracy})
\begin{equation}
\text{\boldmath$\tau$} = \mathbf{M}\times \mathbf{B}_{\mathrm{vac}}
\label{eq:sphtorque}
\end{equation}
which is zero as long as $\mathbf{M}$ and $\mathbf{B}_{\mathrm{vac}}$ are exactly anti-parallel, but will, as soon as $\mathbf{M}$ slightly deviates from being anti-aligned with $\mathbf{B}_{\mathrm{vac}}$, force $\mathbf{M}$ to rotate until it becomes parallel to $\mathbf{B}_{\mathrm{vac}}$. As the spheromak axis is altered, the fixed $\mathbf{B}_{\mathrm{vac}}$ is no-longer appropriately aligned to compensate the Lorentz force $\mathbf{J}\times \mathbf{B}_{\mathrm{bs}}$ acting on $\mathbf{J}$ and the spheromak subsequently ceases to be in equilibrium and disintegrates \citep{mehta_spheromak_tilting_2021}.

The latter does not occur in the case discussed in Section~\ref{ssec:lffsph_surface_currents}, where no background vacuum field is present and surface currents provide the magnetic field contribution that is necessary to balance the Lorentz force acting on the current density inside $V$. In this case, if the spheromak rotates, also the surface currents and the field produced by them will rotate, so that the equilibrium is not destroyed and the force-free Equation~\eqref{eq:forcefreeeq} is also in the rotated state satisfied \citep[][chapter 10.3]{bellan_fundamentals_2000}.\\

\subsection{CME-spheromak and tilting}\label{ssec:cmespheromak_and_tilting}
Which of the above mentioned scenarios applies when using LFF spheromaks to model CMEs that are moving through the ambient solar wind in the inner heliosphere? In the low corona, it might be possible that the tilting instability scenario, where a spheromak is initially anti-aligned to the ambient magnetic field, can occur, as discussed for example in \citet{shiota_mhd_tilting_2010}. But, as the spheromak spends more and more time in the ambient magnetic field while propagating away from the Sun, any inhomogeneity it encounters on its way through the ambient field would quickly trigger the instability, and one can therefore expect that anti-aligned spheromaks should not persist to larger heliocentric distances.
However, as a spheromak that serves as model for a magnetised CME is embedded in a highly-conducting plasma, it does not need a properly anti-aligned background magnetic field to prevent its toroidal current from being accelerated outwards. The outward-pointing Lorentz-force which $\mathbf{B_{\mathrm{bs}}}$ exerts on the current density $\mathbf{J}$ can also be compensated by surface currents as discussed above in Section~\ref{ssec:lffsph_surface_currents}. The latter can, however, not be the full answer to the stability problem, as although the field $\mathbf{B}_{\mathrm{sf}}$ formed by surface currents counteracts the outward pointing Lorentz force due to $\mathbf{B}_{\mathrm{bs}}$ in the interior of the spheromak, the outward pointing Lorentz force which $\mathbf{B}_{\mathrm{bs}}$ and $\mathbf{B}_{\mathrm{sf}}$ exert on the surface current itself is not compensated by any magneto-static force. The real solution lies therefore presumably in the fact that a CME-spheromak is not stationary but expanding, and the Lorentz force produced by $\mathbf{B_{\mathrm{bs}}}$ and (if present) $\mathbf{B_{\mathrm{sf}}}$ can be compensated also by dynamical effects, related to changes in the magnetic flux-density inside the expanding spheromak, as well as by hydrodynamic effects.
In any case, a CME-spheromak might be able to survive a tilting instability in its low-corona evolution. But, regardless of whether such a tilting instability occurs or not, CME-spheromaks that reach a sufficiently large heliocentric distance, can be expected to be well-embedded into their ambient magnetic field and well aligned with it, as they had plenty of time to adapt to the strong field in the low corona. The latter can, however, only occur if the CME-spheromaks indeed had time to evolve through the ambient field. If a CME-spheromak is inserted at higher heliocentric distances, its embedding into the ambient field and its proper alignment has to be ensured manually. The spheromak would then undergo only minor further rotations to maintain the alignment.

It is worth noting that, as the magnetic field of the solar wind is not homogeneous but points at different locations in different directions and becomes quickly weaker with increasing heliocentric distance, a CME-spheromak is in general not only subject to the torque in equation \eqref{eq:sphtorque}, but can also experience a magnetic drift force,
\begin{equation}
\mathbf{F_{\mathrm{drift}}}=\mathbf{\nabla}(\mathbf{M}\cdot\mathbf{B_{\mathrm{vac}}})\stackrel[\mathclap{\overbrace{(\curl(\mathbf{B_{\mathrm{vac}}})=0)}}]{}{\quad=\quad}(\mathbf{M}\cdot\mathbf{\nabla})\mathbf{B_{\mathrm{vac}}}\ ,\label{eq:sphdriftforce}
\end{equation} 
which can affect its trajectory and speed.

Neither of the two static LFF spheromak scenarios discussed above in Sections~\ref{ssec:lffsph_homogeneous_background} and \ref{ssec:lffsph_surface_currents} seems to capture particularly well the overall situation of an expanding CME-spheromak evolving away from the Sun. The details of how such LFF CME-spheromaks interact with the ambient solar wind and ambient magnetic field after having been inserted into the modelling domain of a MHD simulation of the inner heliosphere needs to be better understood and will require further investigation. As a first step in this direction, we present in Section~\ref{sec:results} our findings, that LFF type CME-spheromaks in MHD simulations of the inner heliosphere can be observed to indeed undergo tilting and drifting, and that the dependency of tilting and drifting on input parameters like ambient field strength and orientation, as well as initial spheromak velocity and spheromak orientation, is in qualitative agreement with what one would expect based on the formulas in Equations~\eqref{eq:sphtorque} and \eqref{eq:sphdriftforce} for magnetic torque and magnetic drift force.

\section{Methodology} \label{sec:methods}

\begin{figure}[tb]
\centering
\includegraphics[width=0.66\linewidth,keepaspectratio]{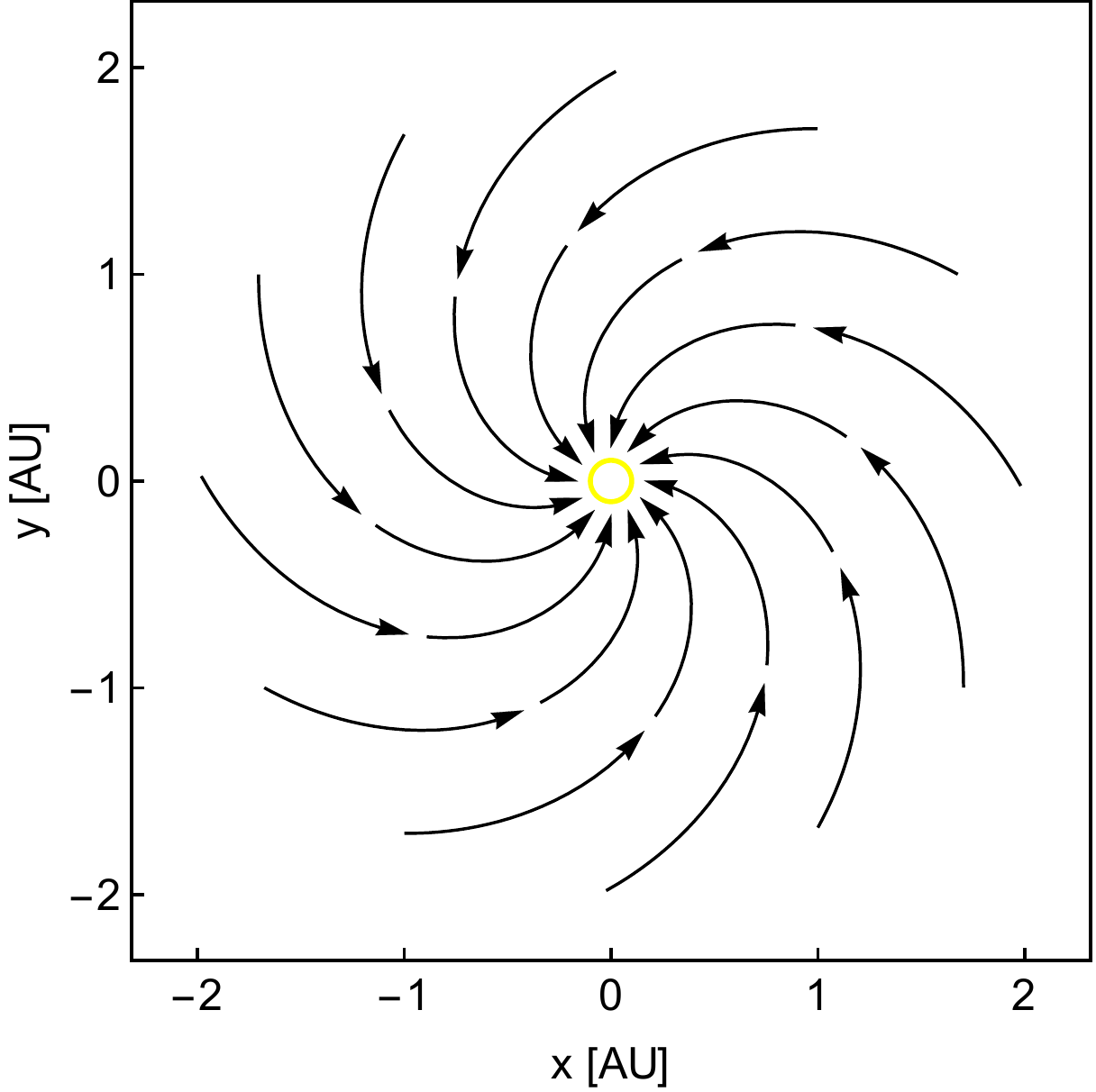}\\[10pt]
\includegraphics[width=0.66\linewidth,keepaspectratio]{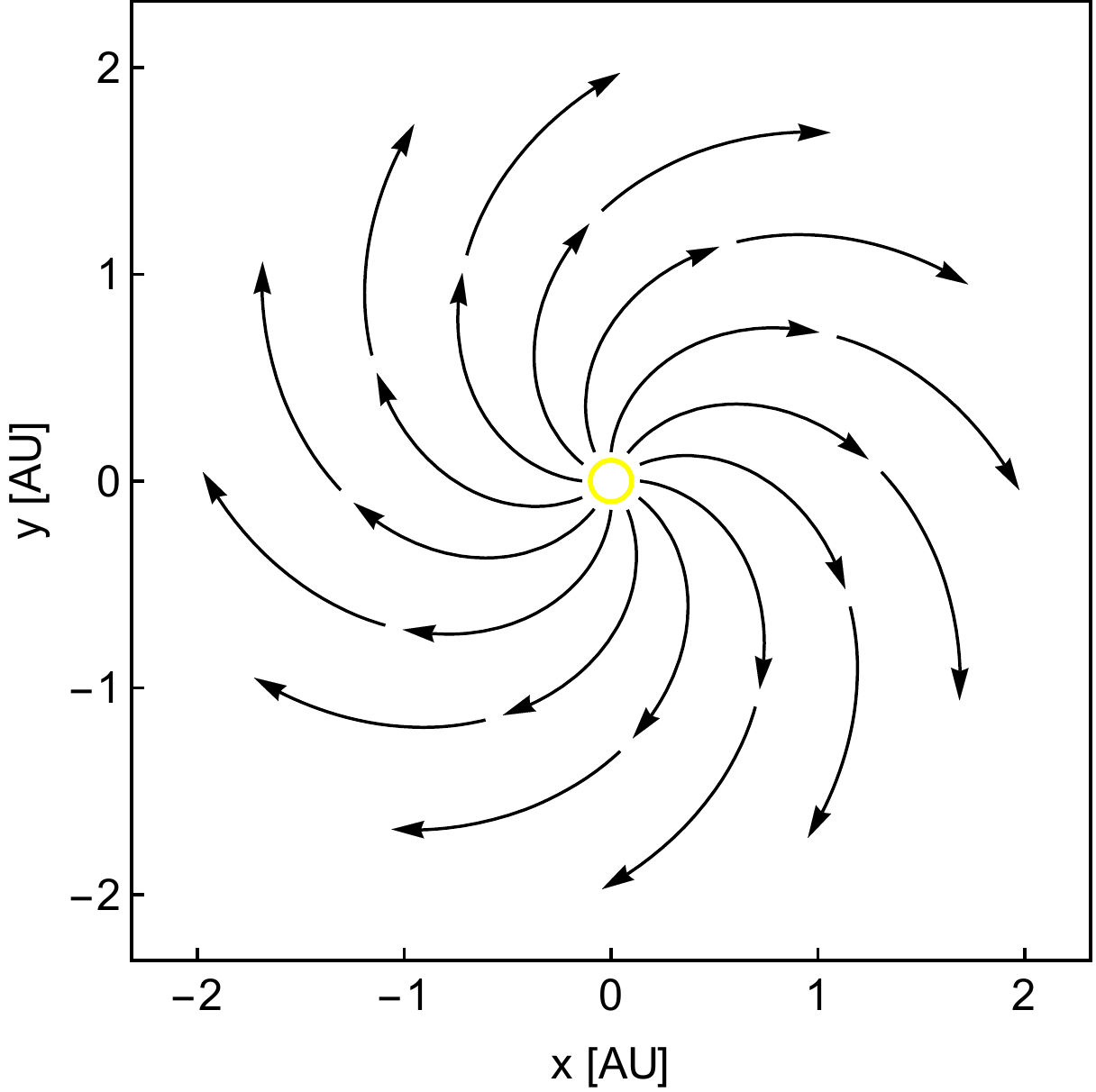}
\caption{Schematics of the magnetic field lines in the equatorial plane of the Sun for the two types of ambient solar wind setups used in this study. In both of these setups, the B-field is set to be predominantly radial at the inner boundary (yellow circle); in the first case, the B-field is inward pointing $B_r<0$ (upper panel), and in the second case outward pointing $B_r>0$ (lower panel).}
\label{fig:parker_spiral}
\end{figure}

For this study we construct four different idealised ambient solar wind plasma and magnetic field conditions, using the EUHFORIA MHD heliospheric model \citep{pomoell_euhforia_2018}, hereafter referred to as background scenarios. In all these background scenarios the ambient magnetic field is defined at the inner radial boundary (set at 0.1 AU) to be predominantly radial (there is a longitudinal magnetic field component due to the Parker spiral) and only its strength and direction (away/towards the Sun) differs. We considered at the inner boundary a weak negative (inward) ($B_{r} = -100~\mathrm{nT}$), a weak positive (outward) ($B_{r} = 100~\mathrm{nT}$), a strong negative ($B_{r} = -300~\mathrm{nT}$), and a strong positive ($B_{r} = 300~\mathrm{nT}$) radial field (see Figure~\ref{fig:parker_spiral} for a simple schematic on the background field orientation in the positive (outward) and negative (inward) cases). From a global point of view, this kind of magnetic field configuration is of course oversimplified, but locally it represents an idealised situation in which a CME is moving along open field lines. 
The inner boundary values for the radial plasma velocity ($v_{r}=450.0~\mathrm{km}/\mathrm{s}$), thermal pressure ($P = 3.3~\mathrm{nPa}$), density ($n = \efnum{5.787}{8}~\mathrm{kg}/\mathrm{m}^{3}$), and temperature ($T = \efnum{4.13}{5}~\mathrm{K}$) were the same for all background scenarios. The MHD simulations were carried out on a spherical grid with 256 radial grid points, covering the distance from 0.1~AU-2.0~AU, and a 4 degree angular resolution for longitude and co-latitude. It is noteworthy that the selected resolution can have an impact on all MHD simulation output, and thus affect for example the concrete numbers obtained for the total rotation angles of the spheromaks. However, it should not change the results present in this paper on a qualitative level, namely that the spheromak experiences a magnetic torque and thus in response undergoes rotation. The degree to which the resolution alters the quantitative results will be discussed in detail in future work.

\begin{deluxetable}{cccc}
\tablenum{1}
\tablecaption{Varying Spheromak input parameters\label{tab:inputparam}}
\tablewidth{0pt}
\tablehead{
\colhead{spheromak No.} & \colhead{speed} & \colhead{insertion tilt} & \colhead{helicity sign}\\
 & \colhead{[$\mathrm{km}/\mathrm{s}$]} & \colhead{[${}^\circ$]} & 
}
\startdata
1 & $900.0$ & 0 & +1  \\
2 & $1350.0$ & 0 & +1  \\
3 & $1800.0$ & 0 & +1 \\
\hline
4 & $900.0$ & 90 & +1 \\
5 & $900.0$ & 180 & +1 \\
6 & $900.0$ & -90 & +1 \\
7 & $900.0$ & 0 & -1 \\
8 & $900.0$ & 90 & -1 \\
9 & $900.0$ & 180 & -1 \\
10 & $900.0$ & -90 & -1 \\
\enddata
\end{deluxetable}

\begin{figure}[htb]
\centering
\includegraphics[width=0.5\linewidth,keepaspectratio]{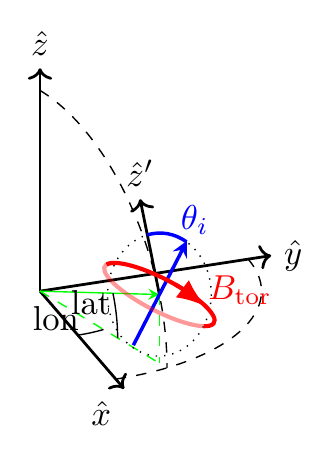}
\caption{Insertion tilt $\theta_i$ for a spheromak that is inserted at the inner boundary at given longitude and latitude $(lon,lat)$. The blue arrow represents the spheromak's axis of symmetry, and points in the direction from which the spheromak's toroidal field (red) would be seen to flow counter-clockwise around its axis. The x--y-plane coincides with the Sun's equatorial plane.}
\label{fig:insertion_tilt}
\end{figure}

In each of the above background scenarios we inserted 10 different types of spheromaks, varying their initial parameters, namely speed, insertion tilt (orientation angle), and helicity sign as shown in table \ref{tab:inputparam}. The employed spheromak model is the EUHFORIA implementation of a LFF spheromak, described in \citet[][]{verbeke_evolution_2019}. In EUHFORIA the insertion tilt of the spheromak is defined as the angle in the tangent plane to the inner boundary, by which the symmetry axis of the spheromak (oriented so that if it is pointing towards us, we see the toroidal magnetic field of the spheromak flowing counter-clockwise) is clock-wise rotated away from the meridional direction, as illustrated in Figure~\ref{fig:insertion_tilt}. The helicity sign is the sign of the parameter $\lambda$ in equation \eqref{eq:forcefreeeq} and can be interpreted as defining whether the current density inside the spheromak is parallel or anti-parallel to the magnetic field. All spheromaks were inserted at the inner boundary at $0^{\circ}$ longitude and latitude, with uniform $\efnum{1.0}{-18}~\mathrm{kg}/\mathrm{m}^{3}$ density, $\efnum{0.8}{6}~\mathrm{K}$ temperature, and a flux content of $\efnum{80.0}{12}~\mathrm{Wb}$. The spheromak radius in all cases was equal to $10~R_{Sun}$. In total we produced 40 unique EUHFORIA simulation runs.

At this point it is worth highlighting that in our analysis we focused on modelled spheromak-CMEs with insertion velocity equals to $v_i=900.0~\mathrm{km}/\mathrm{s}$ or higher. The primary reason for this choice is that fast CMEs are most often of primary concern in operational space weather predictions.   For the slow spheromaks that were tested, namely the ones having insertion velocity equal to that of the ambient plasma ($v_i=450.0~\mathrm{km}/\mathrm{s}$), we observed significant deformations of the spheromaks during the initial phase of the simulations. This effect is due to the combination of spheromak input parameters and the equivalent parameters of the ambient field and plasma. A discussion of the resulting complications in analysing the data would have reduced the clarity of our main results and thus we did not consider slow spheromaks in our analysis. We would like to clarify that a detailed pressure balance assessment needs to be made in order to address this issue. Thus, we caution users of the spheromak model when modelling slow CMEs to carefully select the mass density and flux of the spheromak with respect to the ambient plasma and magnetic field so as to be consistent with the speed of the observed dynamics.

For each combination of background scenario and spheromak, we monitored for the first 70 hours of evolution in the EUHFORIA MHD simulation the position and orientation of the spheromaks every 30 minutes. The used methods for determining the location and orientation of the spheromaks in the simulation output is detailed in Appendix~\ref{apsec:spheromak_monitoring}.

\begin{figure*}[htb]
\centering
\begin{tikzpicture}[scale=0.95,nodes={inner sep=0},every node/.style={transform shape}]
  \node[inner sep=0pt,above right] at (0cm,0cm) {\includegraphics[width=1.0\linewidth,keepaspectratio, trim={2cm 2.5cm 3cm 2.5},clip]{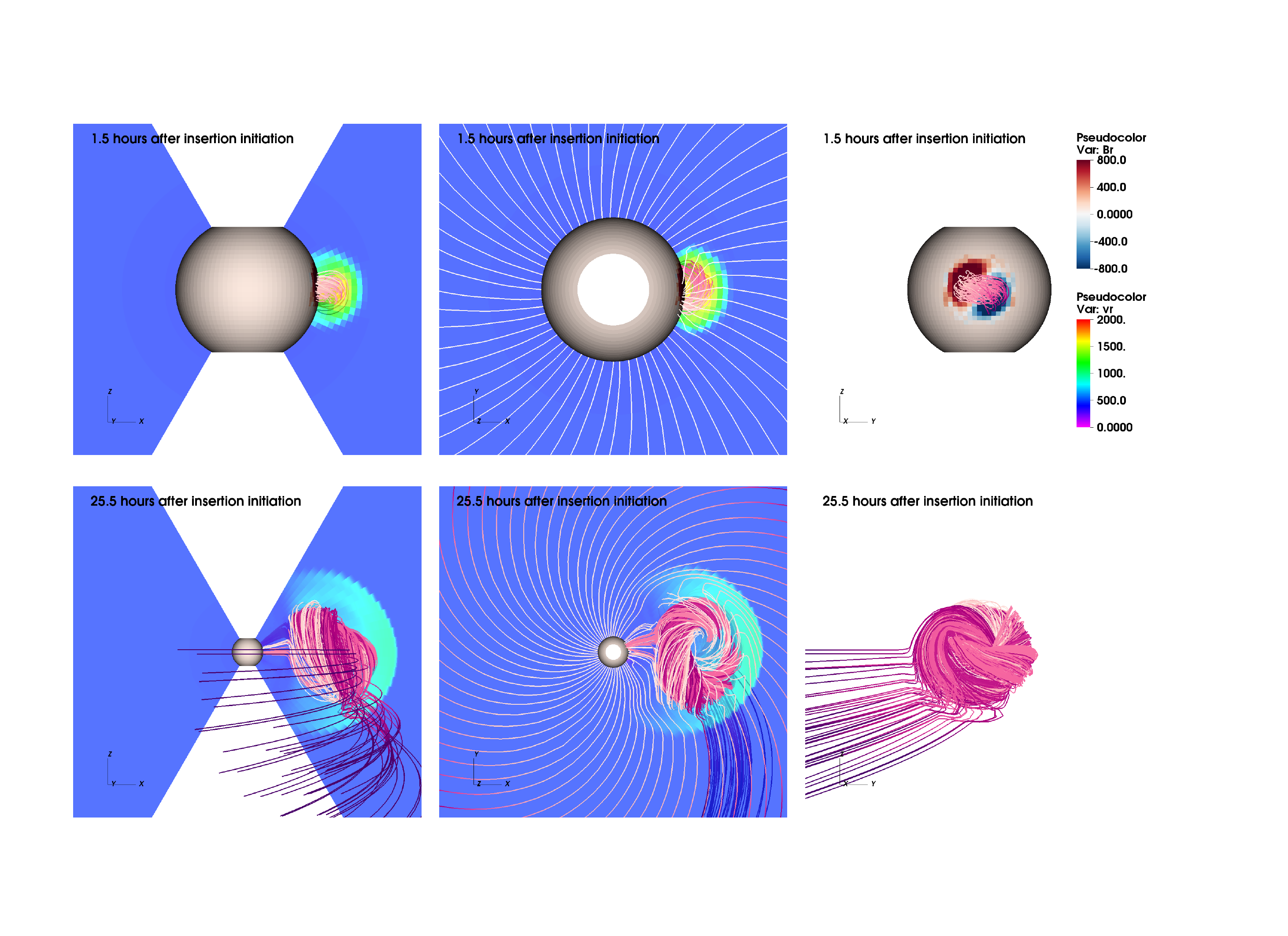}};
  \node (A) at (4.0cm,9.125cm) {};
  \draw[-,line width=0.7mm,black,dashed] ($(A)-(93:0.99cm)$) -- ($(A)+(93:0.99cm)$) node [pos=0.26,rotate=93,anchor=south] {};
  \node (B) at (4.1cm,2.9cm) {};
  \draw[-,line width=0.7mm,black,dashed] ($(B)-(69:1.5cm)$) -- ($(B)+(69:1.5cm)$) node [pos=0.26,rotate=69,anchor=south] {};
\end{tikzpicture}
\caption{Side view (meridional cut given in the left column), top view (equatorial cut given in the middle column), and front view (meridional cut given in the right column) of EUHFORIA MHD simulation output for a spheromak, with initial velocity $v_i=900.0~\mathrm{km}/\mathrm{s}$, insertion tilt $\theta_i={0}^\circ$, and helicity sign $h=1$, 1.5 hours (top row) and 25.5 hours (bottom row) after the insertion initiation. At the inner boundary we plotted the radial magnetic field component, $B_r$. The meridional and equatorial slices in the side and top views (left and middle column respectively) show the spatial distribution of the radial velocity, $v_r$. Both for $B_r$ and $v_r$ the colour maps are the same in all images and are given in the top right image panel. The field line topology of the spheromak is shown in all panels, while the ambient magnetic field lines are plotted on the equatorial plane (middle column panels). The black dashed lines in the left column images (top and bottom row) show the orientation of the magnetic moment of the spheromak.}
\label{fig:euhforia_0tilt}
\end{figure*}
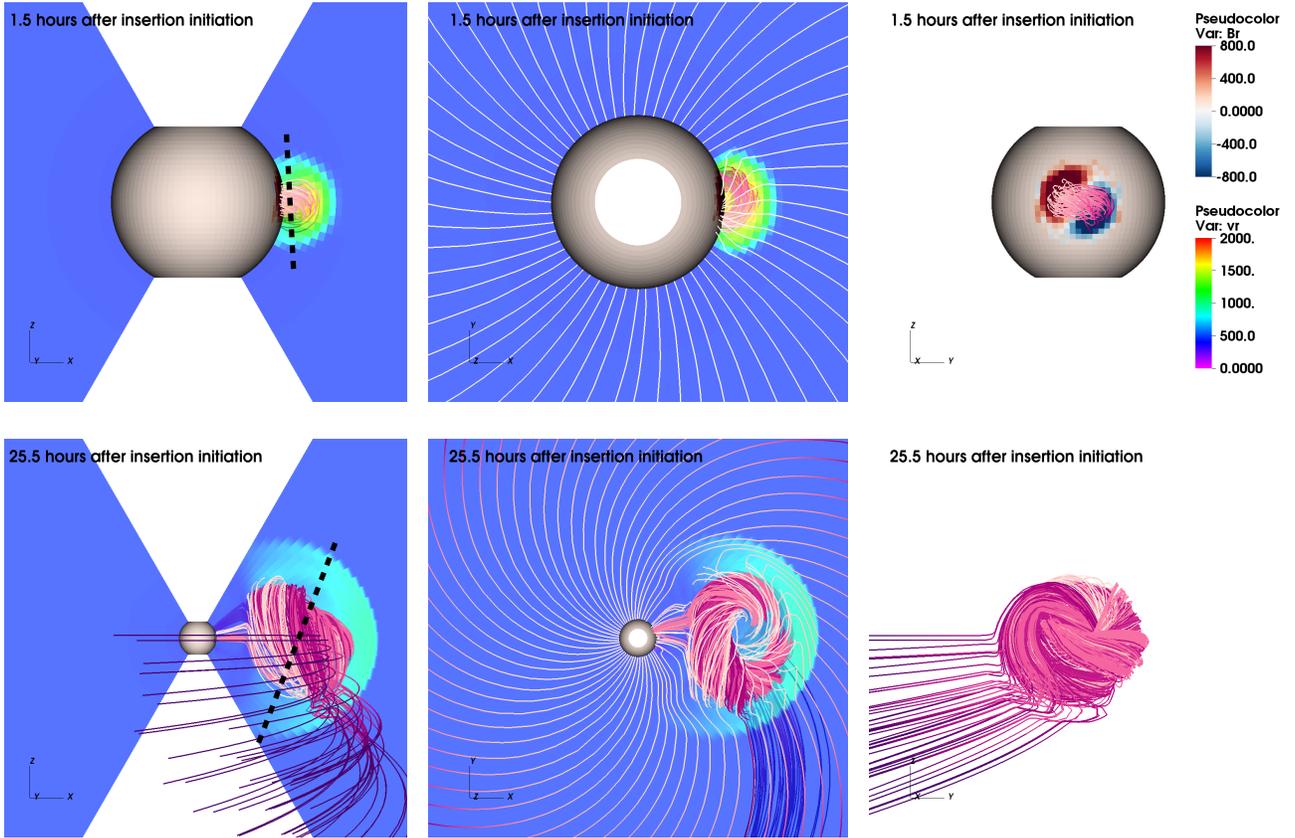

\section{Results}
\label{sec:results}

In Figure~\ref{fig:euhforia_0tilt} we show a visualisation of a selection of magnetic field lines for a spheromak that has entered the EUHFORIA MHD modelling domain through the inner boundary with initial velocity $v_i=900.0~\mathrm{km}/\mathrm{s}$, insertion tilt $\theta_i={0}^\circ$, and helicity sign $h=1$. The ambient magnetic field configuration for the case shown is the weak outward-pointing field described in the previous section. The upper row of the figure shows the side- (left column), top- (middle column), and front-view (right column) of the spheromak, after 1.5 hours from the start of the insertion when the insertion is still in progress. The bottom row shows the system 24 hours later.
Already while the spheromak is being inserted, it appears to experience tilting due to interaction with the ambient field, as can be seen in the top row images, where the spheromak tilt is no longer equal to the insertion tilt value of ${0}^\circ$, already 1.5 hours after the insertion started. Instead the spheromak is tilted slightly backwards (see black dashed line in the top left image). The reason for this initial backward tilting is that the spheromak is still being inserted into the modelling domain (not yet fully out of the inner boundary) and the bottom part of the spheromak gets pushed outwards due to the incoming poloidal field lines getting in conflict with the outward pointing background field. After the spheromak insertion is completed, the spheromak starts rotating to align its magnetic moment with the ambient magnetic field, as expected from the discussion in Section~\ref{sec:spheromakandtilting}, and which is clearly visible 25.5 hours after the start of the insertion (see black dashed line in the bottom left image). Field lines of the background field wrap around the spheromak, and some pass also directly through its centre (bottom middle image). The latter is a further indication for the magnetic moment of the spheromak being already partially aligned with the background field. If a spacecraft was lying at the direction of propagation of the spheromak, it would travel along or near the spheromak symmetry axis and therefore will record strong magnetic flux density but with little rotation of the magnetic field components.\\

In the remainder of this section we will show that an alignment of the spheromak's axis of symmetry -- or more precisely: of its magnetic moment -- with the ambient magnetic field is not specific to the chosen input parameters of the just discussed run, but can be observed quite generally, for all the combinations of background field scenarios and spheromak insertion parameters described in Section~\ref{sec:methods}. In Subsection~\ref{subsec:rotation_and_drift} we discuss how the insertion tilt and insertion velocity of the spheromak affects its total drift and rotation, while in Subsection~\ref{subsec:virtual_spacecraft_time_series} we address the question of how the rotation of the spheromak manifests in in situ measurements of the magnetic field components.

\subsection{Rotation and drift of spheromak}
\label{subsec:rotation_and_drift}
\begin{figure*}[htb]
\centering
\includegraphics[width=0.67\linewidth,keepaspectratio]{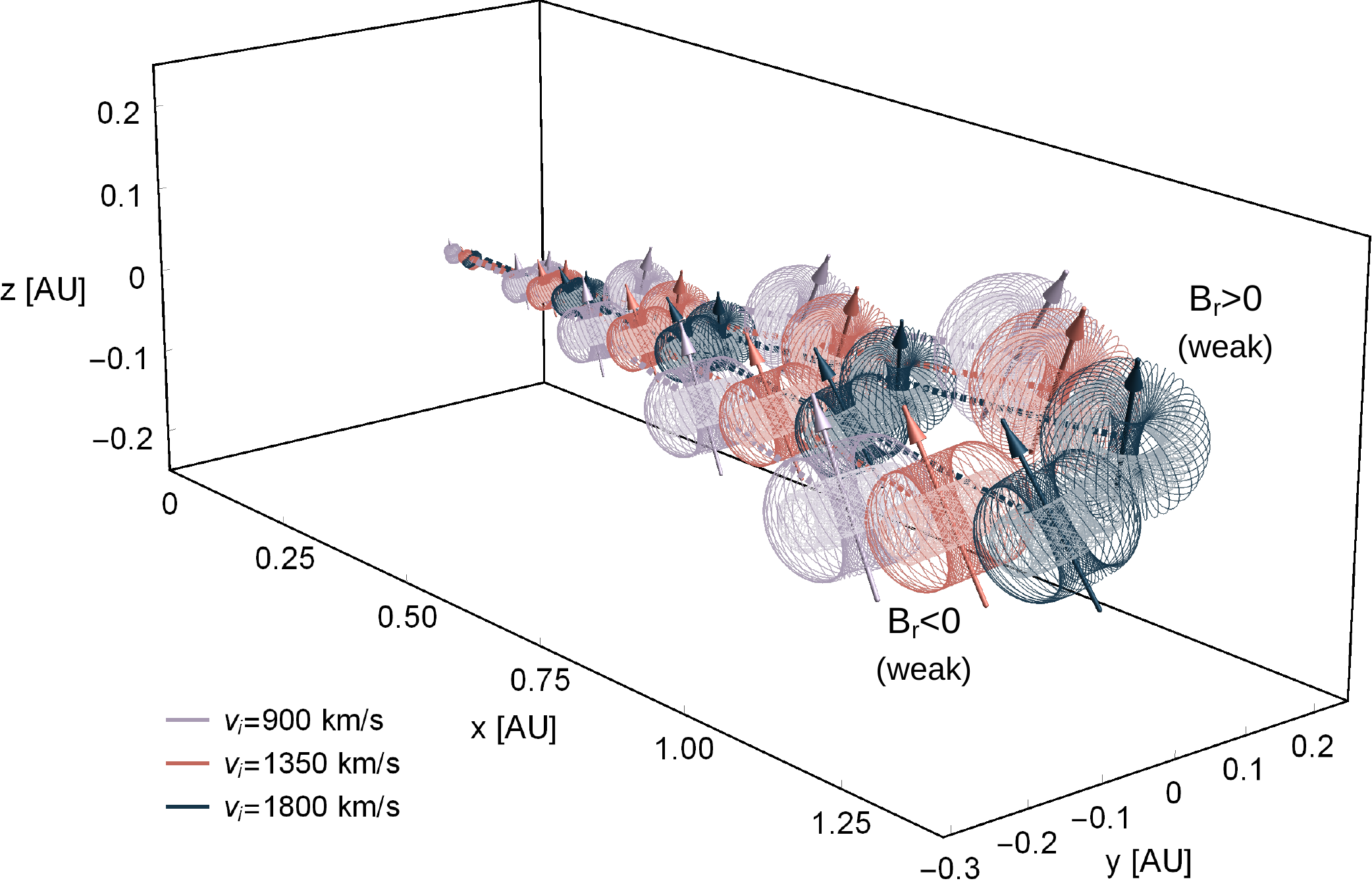}
\includegraphics[width=0.67\linewidth,keepaspectratio]{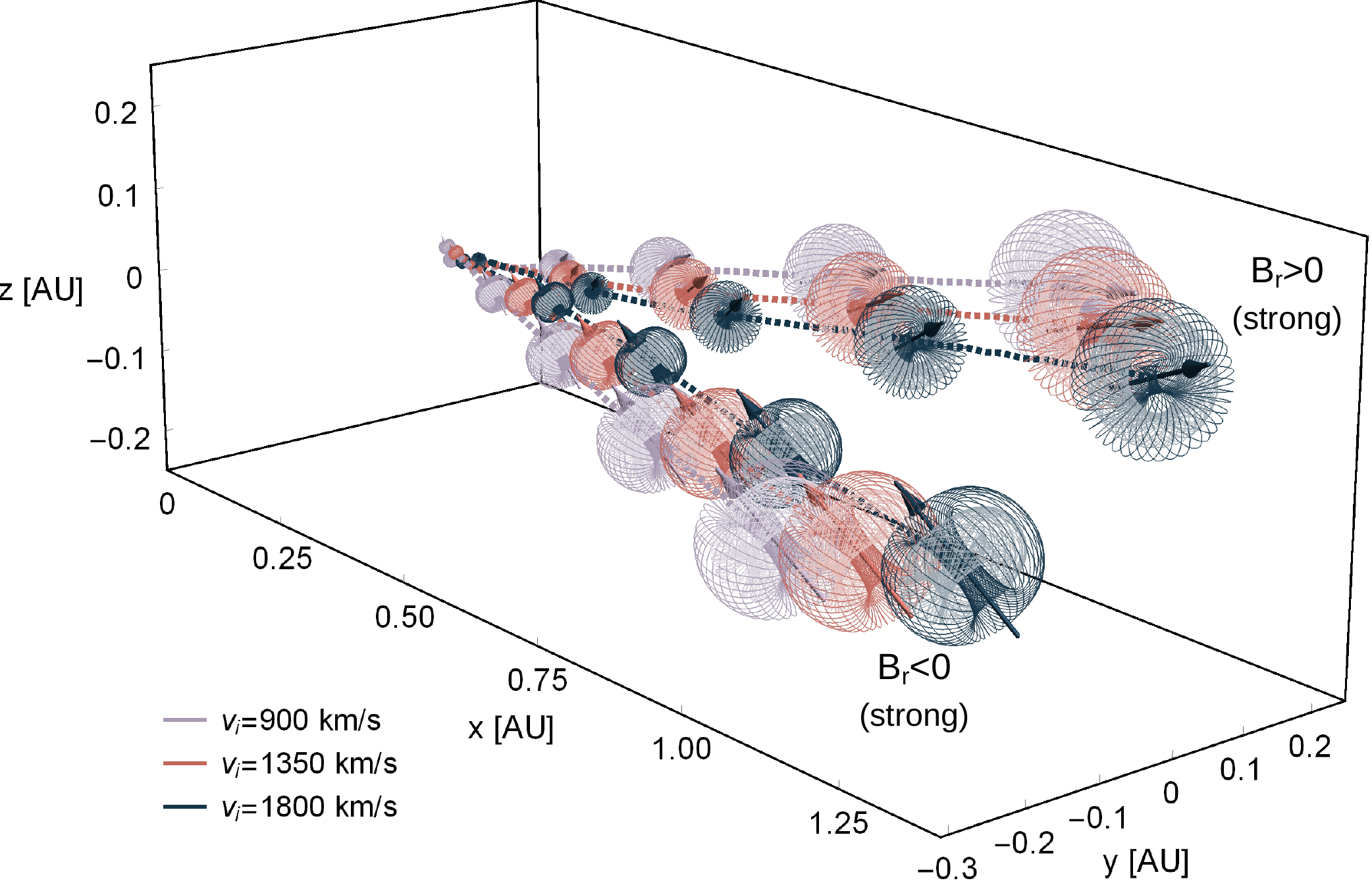}
\caption{Illustration of EUHFORIA simulation results for the evolution of orientation and position of spheromaks with different initial velocity in different ambient field conditions. The shown spheromaks correspond to the cases 1,2 and 3 from Table~\ref{tab:inputparam}: their initial velocity varies, but they all have the same size, field-strength and helicity sign at insertion time, and were inserted at the same location with their magnetic moment pointing in $\mathbf{\hat{z}}$-direction (zero insertion-tilt). Each spheromak-case was inserted in inward- and in outward-pointing background magnetic fields. The magnitude of the ambient magnetic field in the upper figure was $|B_r|=100~\mathrm{nT}$ and in the lower figure $|B_r|=300~\mathrm{nT}$. The evolution of the spheromaks during the EUHFORIA simulation was monitored using the method described in Appendix~\ref{apsec:spheromak_monitoring}. For better visibility, the displayed schematic spheromaks have only 15\% of the size of the corresponding actual spheromaks. The arrows indicate the orientations of the spheromaks' magnetic moments.}
\label{fig:spheromak_drift_and_rot}
\end{figure*}

The background magnetic field at the inner boundary is predominantly radial, and thus the magnetic moments of the inserted spheromaks are initially always at an angle with the background magnetic field. As discussed in Section~\ref{sec:spheromakandtilting}, under such conditions, a spheromak can experience a torque, which acts to align the spheromak's magnetic moment $\mathbf{M}$ with the ambient magnetic field $\mathbf{B_{\mathrm{sw}}}$ of the solar wind, as described in Equation~\eqref{eq:sphtorque}, with $\mathbf{B_{\mathrm{sw}}}$ playing the role of $\mathbf{B_{\mathrm{vac}}}$.
Furthermore, a spheromak can also experiences a magnetic drift force, described by Equation~\eqref{eq:sphdriftforce} and again with $\mathbf{B_{\mathrm{sw}}}$ playing the role of $\mathbf{B_{\mathrm{vac}}}$. Note that this force is strongest if the magnetic moment $\mathbf{M}$ and the ambient magnetic field $\mathbf{B_{\mathrm{sw}}}$ are aligned or anti-aligned.

It should at this point be noted that as the ambient magnetic field varies not just a little bit, but significantly across the volume occupied by a typical CME-spheromak, Equations~\eqref{eq:sphtorque} and \eqref{eq:sphdriftforce} describe the interaction between such a spheromak and the ambient magnetic field only approximately, which is elaborate further in Appendix~\ref{apsec:torque_and_force_accuracy}. Also, it should be stressed that magnetic torque and drift are of course not the only means by which the ambient solar wind and its magnetic field affect the evolution of a CME-spheromak; the spheromak is subject to pressure gradients and other hydrodynamic effects which can interfere with magnetic torque and drift force.

In the remainder of this section, we present our results on tilt and drift of actual CME-spheromaks, evolving in the modelling domain of EUHFORIA MHD simulations.  With the methods described in Appendix~\ref{apsec:spheromak_monitoring} we monitored the location and orientation of spheromaks as they evolve through different ambient solar wind scenarios. We then analysed the dependency of the spheromak's time-evolution on their insertion velocities $v_i$, insertion tilts $\theta_i$, and helicity signs $h$.

\subsubsection{Spheromaks of different initial speeds}
\label{subsubsec:diff_initial_speeds}
\begin{figure*}[htb]
\centering
\includegraphics[width=0.9\linewidth,keepaspectratio]{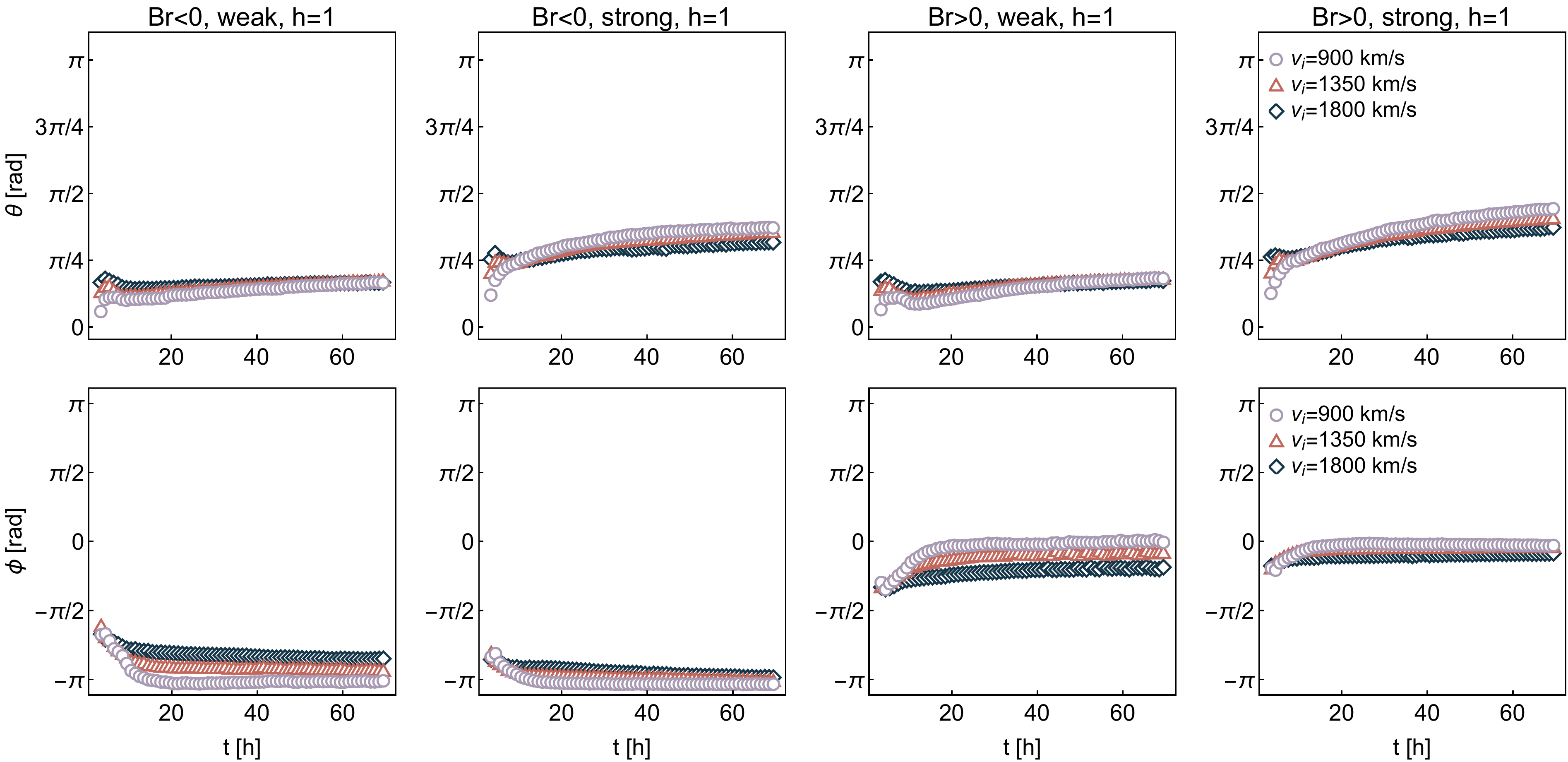}
\caption{EUHFORIA simulation results showing the evolution of the orientation of the magnetic moment of a spheromak for different initial velocities, $v_{i}\in\{900.0~\mathrm{km}/\mathrm{s},1350.0~\mathrm{km}/\mathrm{s},1800.0~\mathrm{km}/\mathrm{s}\}$, and different ambient field conditions. All spheromaks were inserted with insertion-tilt $\theta_i=0$~rad and helicity sign $h=1$ (cases 1-3 from Table~\ref{tab:inputparam}). The orientation of the spheromak's magnetic moment is encoded in the $\theta$ and $\phi$ angles as described in Appendix~\ref{apsec:coordinate_transform}. The two rows show, respectively, the time-evolution of $\theta$ and $\phi$ in a weak ambient field with negative $B_r$ (first column), a strong ambient field with negative $B_r$ (second column) a weak ambient field with positive $B_r$ (third column) and a strong ambient field with positive $B_r$ (last column).}
\label{fig:rotation_vs_t_var_initvel}
\end{figure*}

Figure~\ref{fig:spheromak_drift_and_rot} provides a visual overview of our simulation results, showing how the evolution of a spheromak's magnetic centre of mass and orientation of magnetic moment in different ambient field configurations with $B_r=\pm 100~\mathrm{nT}$ (upper panel) and $B_r=\pm 300~\mathrm{nT}$ (lower panel), depends on the insertion velocity, $v_i=900.0~\mathrm{km}/\mathrm{s}$, $1350.0~\mathrm{km}/\mathrm{s}$,
 and $1800.0~\mathrm{km}/\mathrm{s}$, of the spheromak. All displayed spheromaks had a positive helicity sign, were inserted with zero insertion-tilt (spheromaks number 1-3 in Table~\ref{tab:inputparam}), and were initially propagating along the $x$-axis.\\
The figure indicates, that the spheromaks indeed must have experienced drift-acceleration, which caused their trajectories to deviate from the $x$-axis.  Also, a rotation of the spheromaks is clearly visible, since with increasing distance, the magnetic moments become increasingly aligned with the background magnetic field.
As according to equations \eqref{eq:sphtorque} and \eqref{eq:sphdriftforce}, the approximate torque and drift forces acting on a spheromak are proportional to the magnitude of the background magnetic field, $\mathbf{B_{\mathrm{sw}}}$, spheromaks rotate and accelerate slower in a weak ambient field (Figure~\ref{fig:spheromak_drift_and_rot} upper panel) than in a stronger one (Figure~\ref{fig:spheromak_drift_and_rot} lower panel).
Furthermore, as the ambient magnetic field gets weaker with increasing distance from the Sun, slow spheromaks spend more time in a stronger ambient magnetic field than faster spheromaks, and therefore experience a larger amount of rotation and drift when subject to the same ambient field conditions.

Figure~\ref{fig:rotation_vs_t_var_initvel} shows quantitatively the evolution of the orientation of the magnetic moments of the spheromaks as visually depicted in Figure~\ref{fig:spheromak_drift_and_rot}.
In this figure, the orientation of the magnetic moment of a spheromak is parametrised by means of a pair of polar angles $(\theta,\phi)$, where $\phi \in (-\pi,\pi]$ describes the angle between the coordinate x-direction and the projection of the magnetic moment into the x-y-plane, and $\theta\in [0,\pi]$ is the angle between the magnetic moment and the coordinate z-direction (see Appendix~\ref{apsec:coordinate_transform} and Figure~\ref{fig:coord_transform}). 

Table~\ref{tab:totrotanglesvsvi} lists the total change in rotation and in y- and z-position of spheromaks as they travel from their insertion point at the inner boundary at 0.1~AU to a distance of 1.0~AU. It is apparent that spheromaks inserted in stronger background magnetic field experience stronger total rotation and transverse drift. In addition, the total rotation supports the conclusions made based on the trends in Figures~\ref{fig:spheromak_drift_and_rot} and \ref{fig:rotation_vs_t_var_initvel}. Namely, that faster spheromaks achieve smaller total rotation angles and drift distances, as they reach the larger heliodistances more quickly, where the ambient field is weakened. For the weaker ambient field scenarios, this effect is much smaller and seems to approach the accuracy limits of our spheromak monitoring method.   

\begin{deluxetable}{rccccc}
\tablenum{2}
\tablecaption{Total change in rotation and in y- and z-position for a spheromak with helicity sign $h=1$, insertion tilt $\theta_i=0^{\circ}$, and different insertion velocities, $v_i=900,\,1350,\,1800~\mathrm{km}/\mathrm{s}$, as it travels through different ambient fields from 0.1~AU to 1.0~AU heliocentric distance.\label{tab:totrotanglesvsvi}}
\tablewidth{0pt}
\tablehead{
\colhead{ambient field} & \colhead{ins. vel.} & \colhead{tot. rot.} & \colhead{$\Delta y$} & \colhead{$\Delta z$}\\
 & \colhead{[$\mathrm{km}/\mathrm{s}$]} & \colhead{[${}^\circ$]} & \colhead{[AU]} & \colhead{[AU]}
}
\startdata
 weak $B_r<0$ & 900 & 29.0 & -0.13 & -0.041 \\
 weak $B_r<0$ & 1350 & 28.6 & -0.079 & -0.024 \\
 weak $B_r<0$ & 1800 & 28.7 & -0.054 & -0.002 \\
 \hline
 weak $B_r>0$ & 900 & 31.9 & 0.066 & 0.075 \\
 weak $B_r>0$ & 1350 & 30.0 & 0.022 & 0.060 \\
 weak $B_r>0$ & 1800 & 29.0 & 0.002 & 0.042 \\
 \hline
 strong $B_r<0$ & 900 & 66.0 & -0.208 & -0.063 \\
 strong $B_r<0$ & 1350 & 61.5 & -0.158 & -0.053 \\
 strong $B_r<0$ & 1800 & 54.5 & -0.118 & -0.035 \\
 \hline
 strong $B_r>0$ & 900 & 77.7 & 0.129 & 0.128 \\
 strong $B_r>0$ & 1350 & 69.7 & 0.084 & 0.103 \\
 strong $B_r>0$ & 1800 & 63.0 & 0.051 & 0.076 \\
\enddata
\end{deluxetable}

\subsubsection{Spheromaks with different insertion tilts}
\label{subsubsec:diff_insertion_tilts}
\begin{figure*}[htb]
\centering
\includegraphics[width=0.9\linewidth,keepaspectratio]{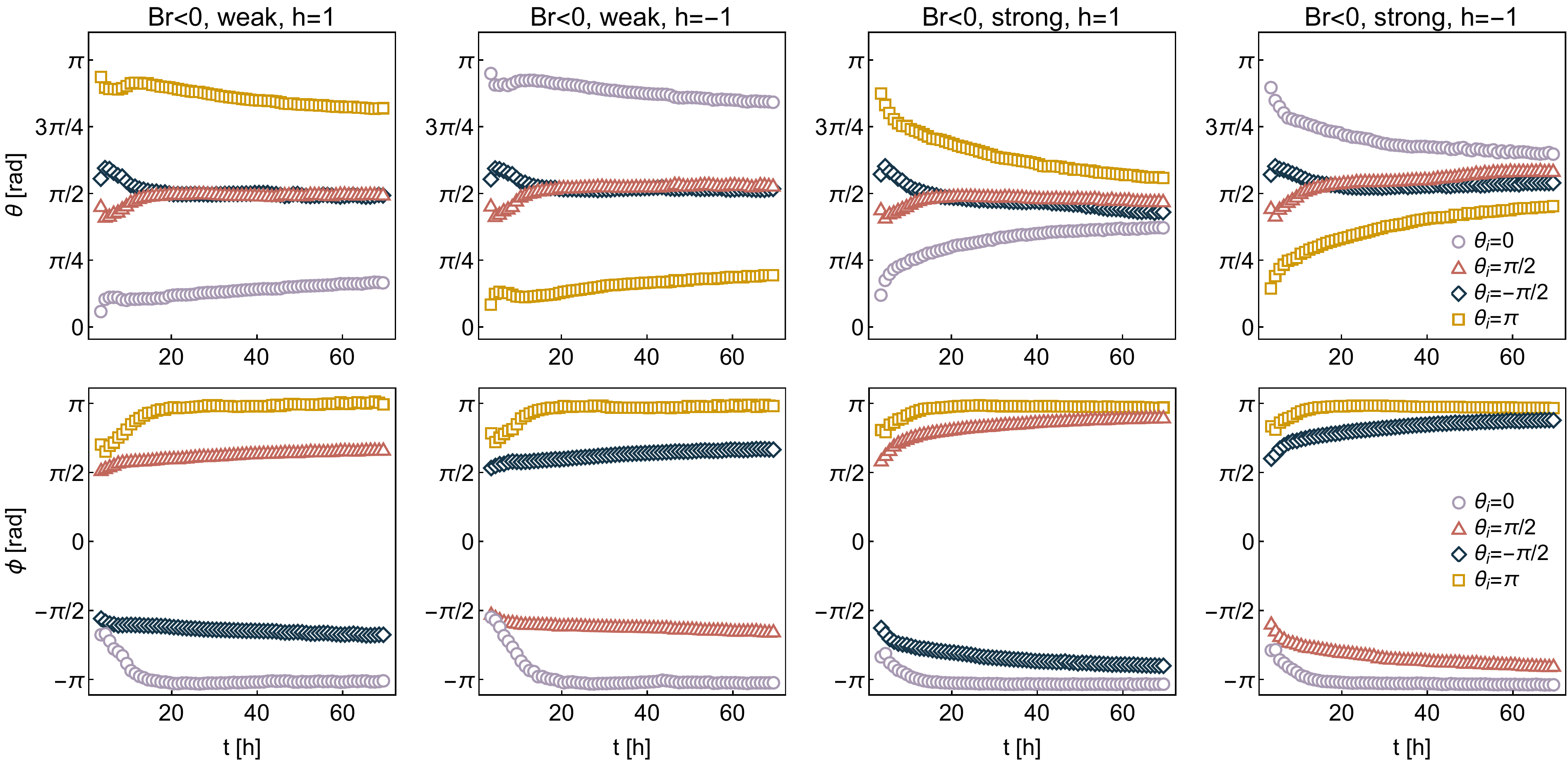}\\[10pt]
\includegraphics[width=0.9\linewidth,keepaspectratio]{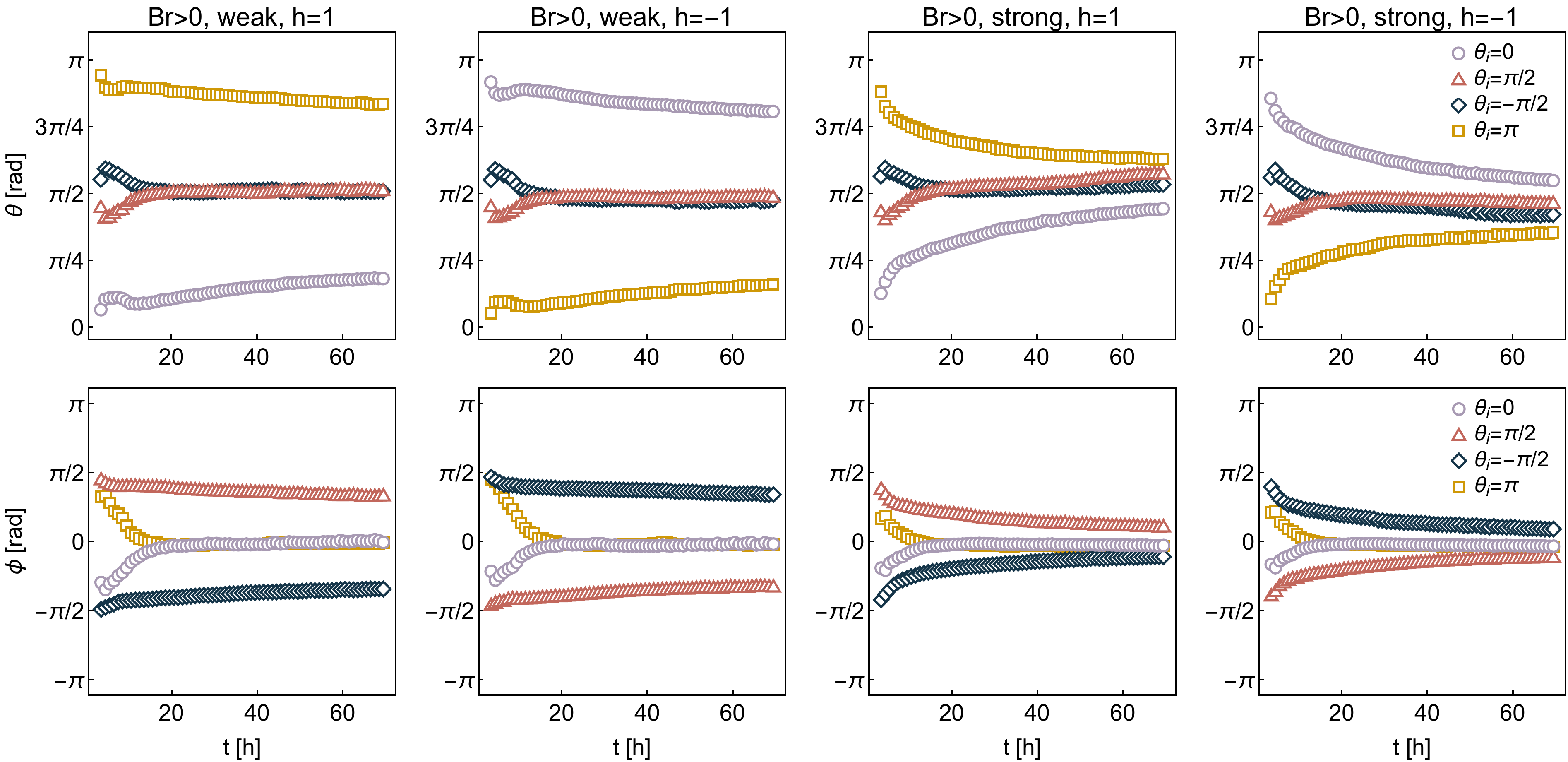}
\caption{EUHFORIA simulation results showing the evolution of the orientation of the magnetic moment of a spheromak for different insertion-tilt angles, $\theta_{i}\in\{0,\pi/2,-\pi/2,\pi\}$, and different ambient field conditions. All spheromaks were inserted with initial speed $v_i=900.0~\mathrm{km}/\mathrm{s}$ and helicity sign $h=\pm 1$ (cases 1 and 4-10 from Table~\ref{tab:inputparam}). The orientation of the spheromak's magnetic moment is encoded in the $\theta$ and $\phi$ angles as described in Appendix~\ref{apsec:coordinate_transform}. The first two rows show, respectively, the time-evolution of $\theta$ and $\phi$ for the four insertion-tilt angles in a weak (first two columns) and strong (last two columns) ambient magnetic field with negative $B_r$, while the last two rows show the corresponding situations for weak and strong ambient field with positive $B_r$.}
\label{fig:rotation_vs_t_var_tilt}
\end{figure*}

Figure~\ref{fig:rotation_vs_t_var_tilt} provides a comparison of simulation results for the time-evolution of spheromaks that have been inserted into the four background field scenarios ($B_r<0$ weak, $B_r<0$ strong, $B_r>0$ weak, and $B_r>0$ strong) with the same initial velocity, $v_i=900.0~\mathrm{km}/\mathrm{s}$, but different insertion tilts, $\theta_i\in\{0,\pi/2,-\pi/2,\pi\}$, and different helicities $h=\pm1$. The data shows that regardless of the insertion tilt, the spheromak's magnetic moment rotates to align itself with the background magnetic field. In the case of weak inward and outward pointing background fields (first two columns of the figure), the corresponding torques acting on the spheromaks are relatively weak and the spheromaks therefore rotate only slowly. In the case of the stronger background fields, the spheromaks rotate much faster, so that their magnetic moments reach almost perfect alignment with the background magnetic field towards the end of the monitored 70 hours simulation time interval.
Table~\ref{tab:totrotanglesvsthi} lists the total change in rotation and in y- and z-position of the spheromaks as they travel from their insertion point at the inner boundary at 0.1~AU to a distance of 1.0~AU. As the ideal-MHD evolution equations are invariant with respect to a global sign-flip of the magnetic field, a simultaneous change of sign of the ambient magnetic field and flip of the spheromak's insertion tilt by $180^{\circ}$ leaves the total rotation angle and total transverse drift unchanged. The input parameters ("weak $B_r<0$", $h=1$, $\theta_i=0^{\circ}$) result therefore in the same total rotation angle and transverse drift as ("weak $B_r>0$", $h=1$, $\theta_i=180^{\circ}$).

\begin{deluxetable}{rccccc}
\tablenum{3}
\tablecaption{Total change in rotation and in y- and z-position as a spheromak that was inserted into different ambient fields with velocity $v_i=900.0~\mathrm{km}/\mathrm{s}$ but with different helicity signs, h, and different insertion tilt values travels from 0.1~AU to 1.0~AU heliocentric distance.}\label{tab:totrotanglesvsthi}
\tablewidth{0pt}
\tablehead{
\colhead{ambient field} & \colhead{h} & \colhead{ins. tilt} & \colhead{tot. rot.} & \colhead{$\Delta y$} & \colhead{$\Delta z$}\\
 & & \colhead{[${}^{\circ}$]} & \colhead{[${}^\circ$]} & \colhead{[AU]} & \colhead{[AU]}
}
\startdata
 weak $B_r<0$ & 1 & 0 & 29.0 & -0.13 & -0.041 \\
 weak $B_r<0$ & 1 & 90 & 26.9 & -0.08 & 0.121 \\
 weak $B_r<0$ & 1 & -90 & 30.5 & 0.03 & -0.076 \\
 weak $B_r<0$ & 1 & 180 & 31.9 & 0.066 & 0.075 \\
 weak $B_r<0$ & -1 & 0 & 28.2 & -0.127 & 0.079 \\
 weak $B_r<0$ & -1 & 90 & 28.3 & 0.032 & 0.124 \\
 weak $B_r<0$ & -1 & -90 & 29.4 & -0.078 & -0.071 \\
 weak $B_r<0$ & -1 & 180 & 34.1 & 0.068 & -0.038 \\
 \hline
 weak $B_r>0$ & 1 & 0 & 31.9 & 0.066 & 0.075 \\
 weak $B_r>0$ & 1 & 90 & 30.5 & 0.03 & -0.076 \\
 weak $B_r>0$ & 1 & -90 & 26.9 & -0.08 & 0.121 \\
 weak $B_r>0$ & 1 & 180 & 29.0 & -0.13 & -0.041 \\
 weak $B_r>0$ & -1 & 0 & 34.1 & 0.068 & -0.038 \\
 weak $B_r>0$ & -1 & 90 & 29.4 & -0.078 & -0.071 \\
 weak $B_r>0$ & -1 & -90 & 28.3 & 0.032 & 0.124 \\
 weak $B_r>0$ & -1 & 180 & 28.2 & -0.127 & 0.079 \\
 \hline
 strong $B_r<0$ & 1 & 0 & 66.0 & -0.208 & -0.063 \\
 strong $B_r<0$ & 1 & 90 & 69.4 & -0.131 & 0.205 \\
 strong $B_r<0$ & 1 & -90 & 71.2 & 0.043 & -0.151 \\
 strong $B_r<0$ & 1 & 180 & 77.7 & 0.129 & 0.128 \\
 strong $B_r<0$ & -1 & 0 & 62.2 & -0.208 & 0.105 \\
 strong $B_r<0$ & -1 & 90 & 72.9 & 0.043 & 0.196 \\
 strong $B_r<0$ & -1 & -90 & 67.8 & -0.132 & -0.158 \\
 strong $B_r<0$ & -1 & 180 & 79.9 & 0.128 & -0.085 \\
 \hline
 strong $B_r>0$ & 1 & 0 & 77.7 & 0.129 & 0.128 \\
 strong $B_r>0$ & 1 & 90 & 71.2 & 0.043 & -0.151 \\
 strong $B_r>0$ & 1 & -90 & 69.4 & -0.131 & 0.205 \\
 strong $B_r>0$ & 1 & 180 & 66.0 & -0.208 & -0.063 \\
 strong $B_r>0$ & -1 & 0 & 79.9 & 0.128 & -0.085 \\
 strong $B_r>0$ & -1 & 90 & 67.8 & -0.132 & -0.158 \\
 strong $B_r>0$ & -1 & -90 & 72.9 & 0.043 & 0.196 \\
 strong $B_r>0$ & -1 & 180 & 62.2 & -0.208 & 0.105 \\
 \enddata
 \end{deluxetable}

\subsection{Rotation signatures in situ at virtual spacecraft}
\label{subsec:virtual_spacecraft_time_series}

\begin{figure*}[htb]
\centering
\includegraphics[height=0.7\linewidth,keepaspectratio]{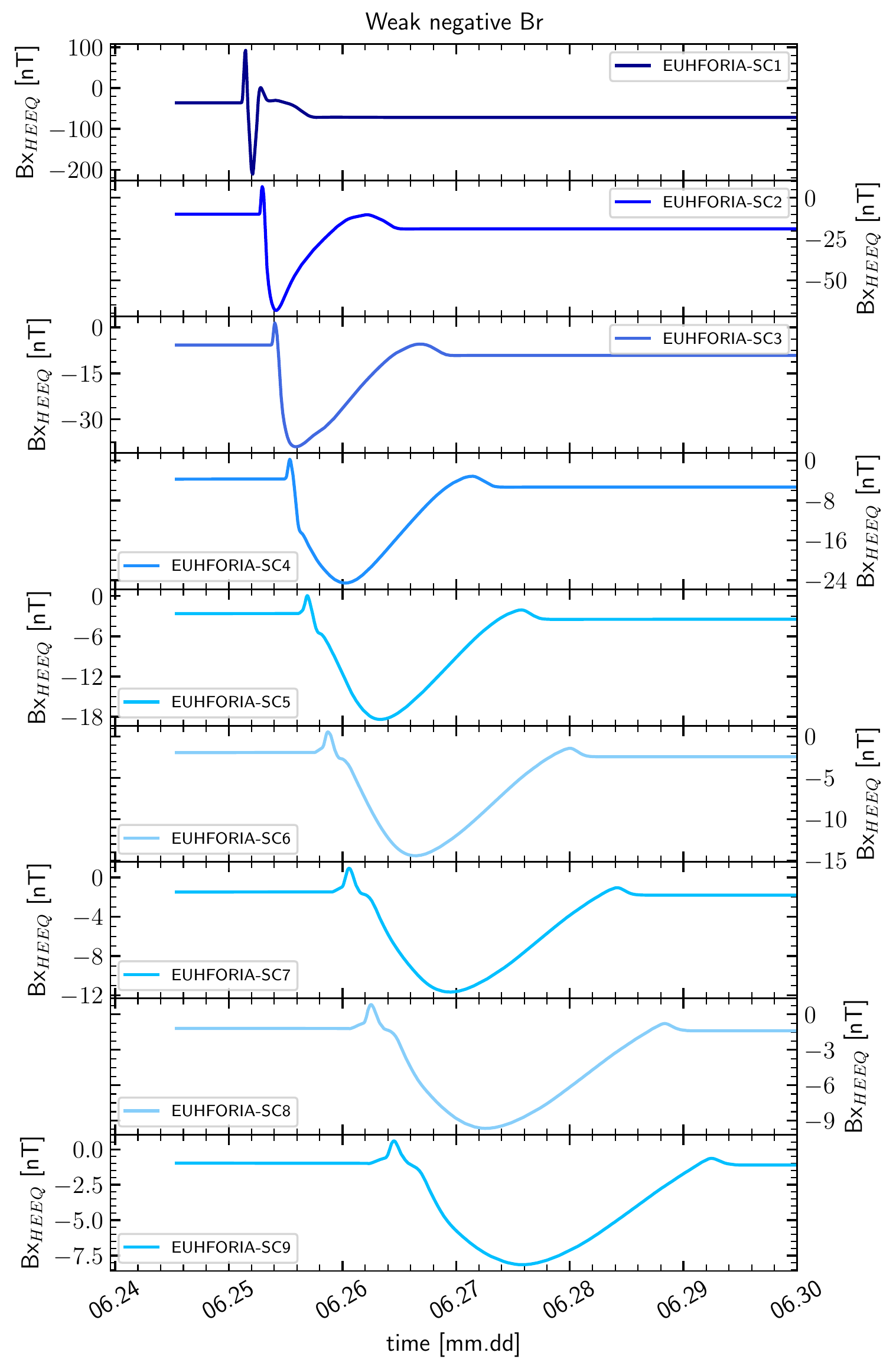}
\includegraphics[height=0.7\linewidth,keepaspectratio]{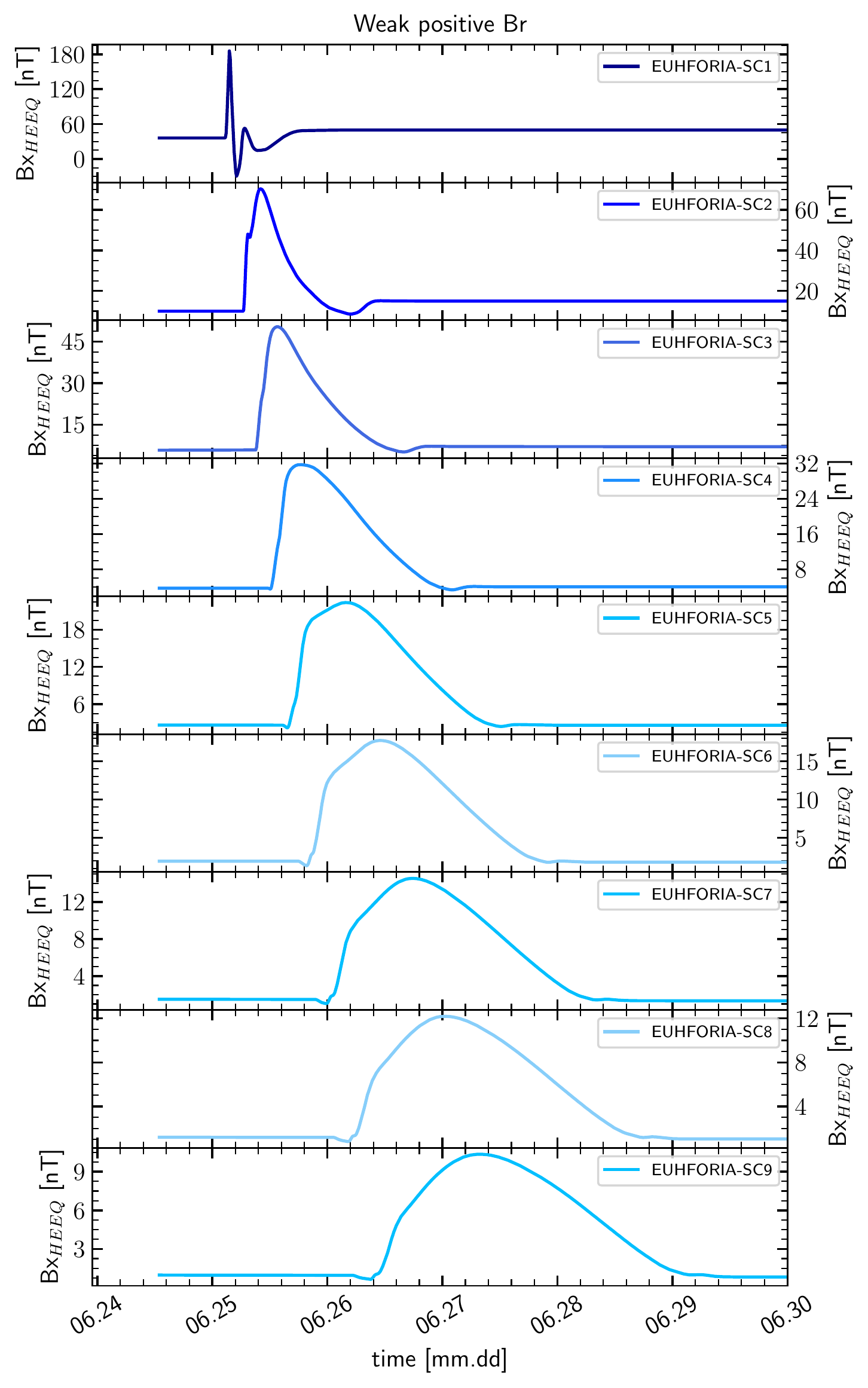}
\caption{Time series of the $B_x$ component in the HEEQ of the magnetic field of the spheromak, inserted with initial velocity $v_i=900.0~\mathrm{km}/\mathrm{s}$, at different virtual spacecraft locations along the Sun-Earth line (SC1: 0.15AU, SC2: 0.3AU, SC3: 0.4AU, SC4: 0.5AU, SC5: 0.6AU, SC6: 0.7AU, SC7: 0.8AU, SC8: 0.9AU, and SC9: 1.0AU heliocentric distance). The left image shows the time series for the case of the spheromak being inserted into the weak negative $B_r$ background magnetic field scenario, while the right image is for the weak positive $B_r$ background field scenario. As one can see, the spheromak's $B_x$ component adapts, due to rotation of the spheromak, to the sign of the ambient magnetic field to which the spheromak is inserted: negative when inserted in a negative $B_r$ ambient field and positive when inserted in a positive $B_r$ ambient field.}
\label{fig:vr_time_series_xweak}
\end{figure*}

\begin{figure*}[htb]
\centering
\includegraphics[height=0.7\linewidth,keepaspectratio]{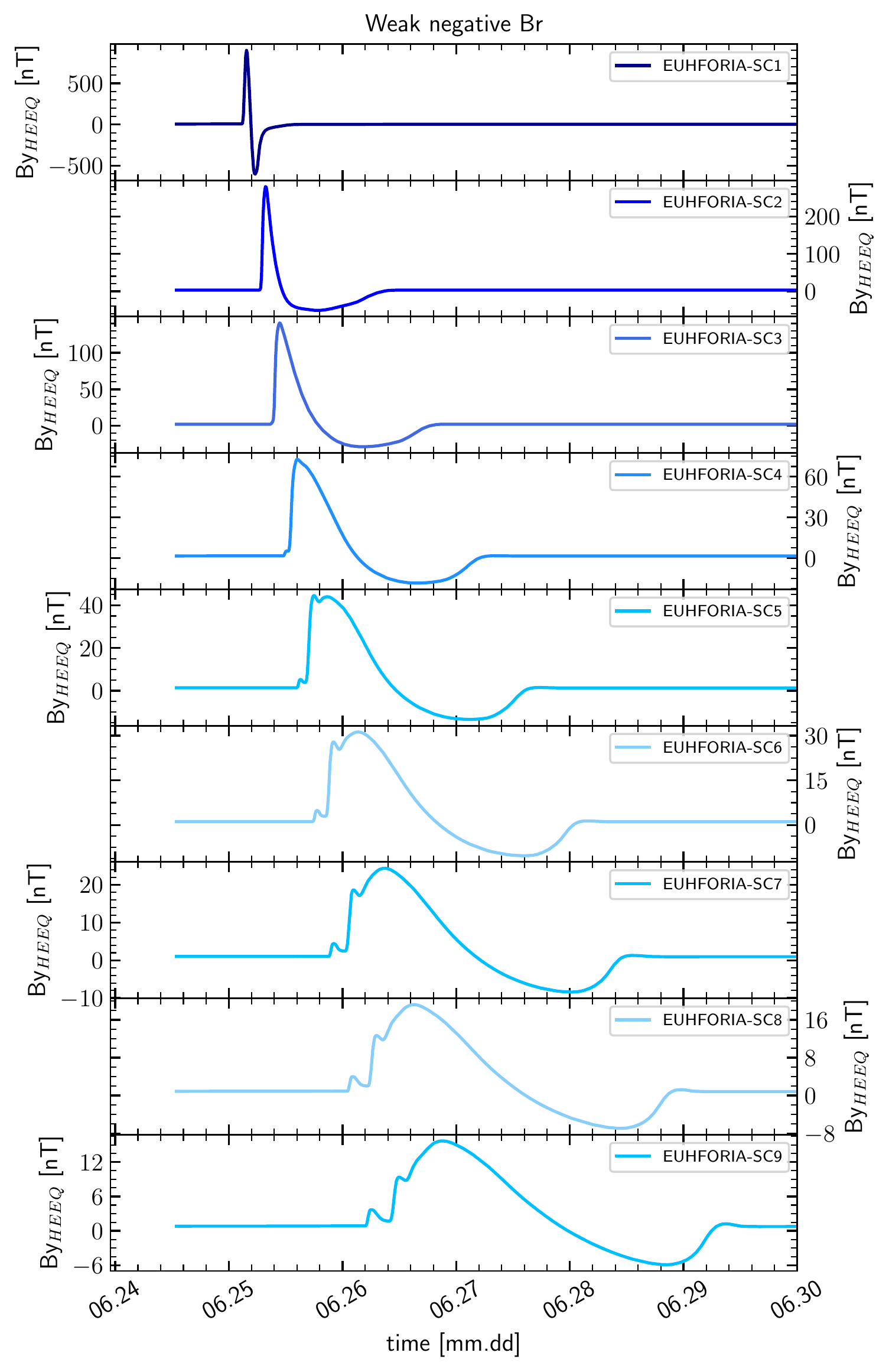}
\includegraphics[height=0.7\linewidth,keepaspectratio]{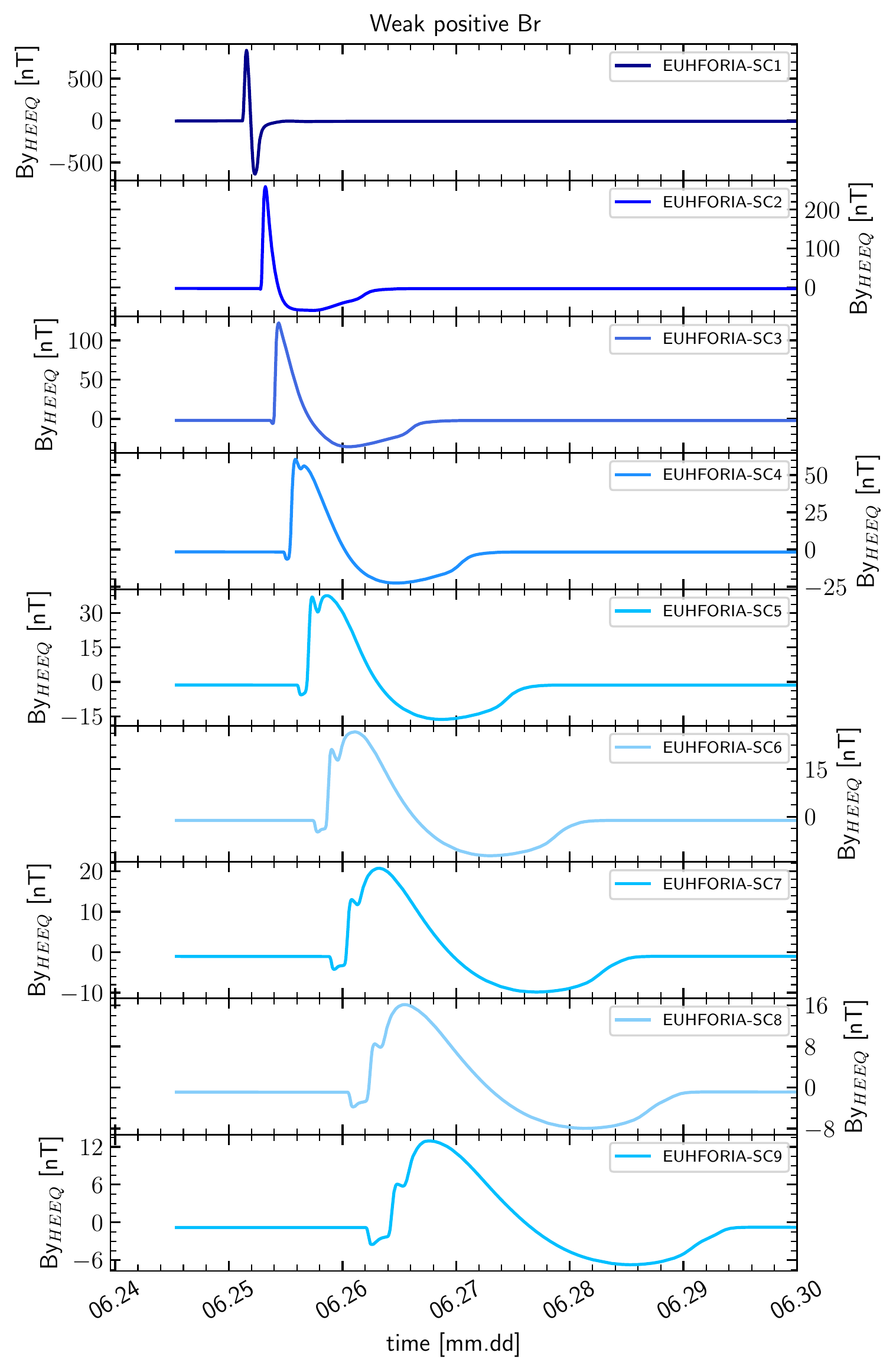}
\caption{Same as Figure \ref{fig:vr_time_series_xweak} but for the $B_y$ component in HEEQ of a spheromak inserted in a weak (left: negative, right:positive) predominantly radial ambient magnetic field.}
\label{fig:vr_time_series_yweak}
\end{figure*}

\begin{figure*}[htb]
\centering
\includegraphics[height=0.7\linewidth,keepaspectratio]{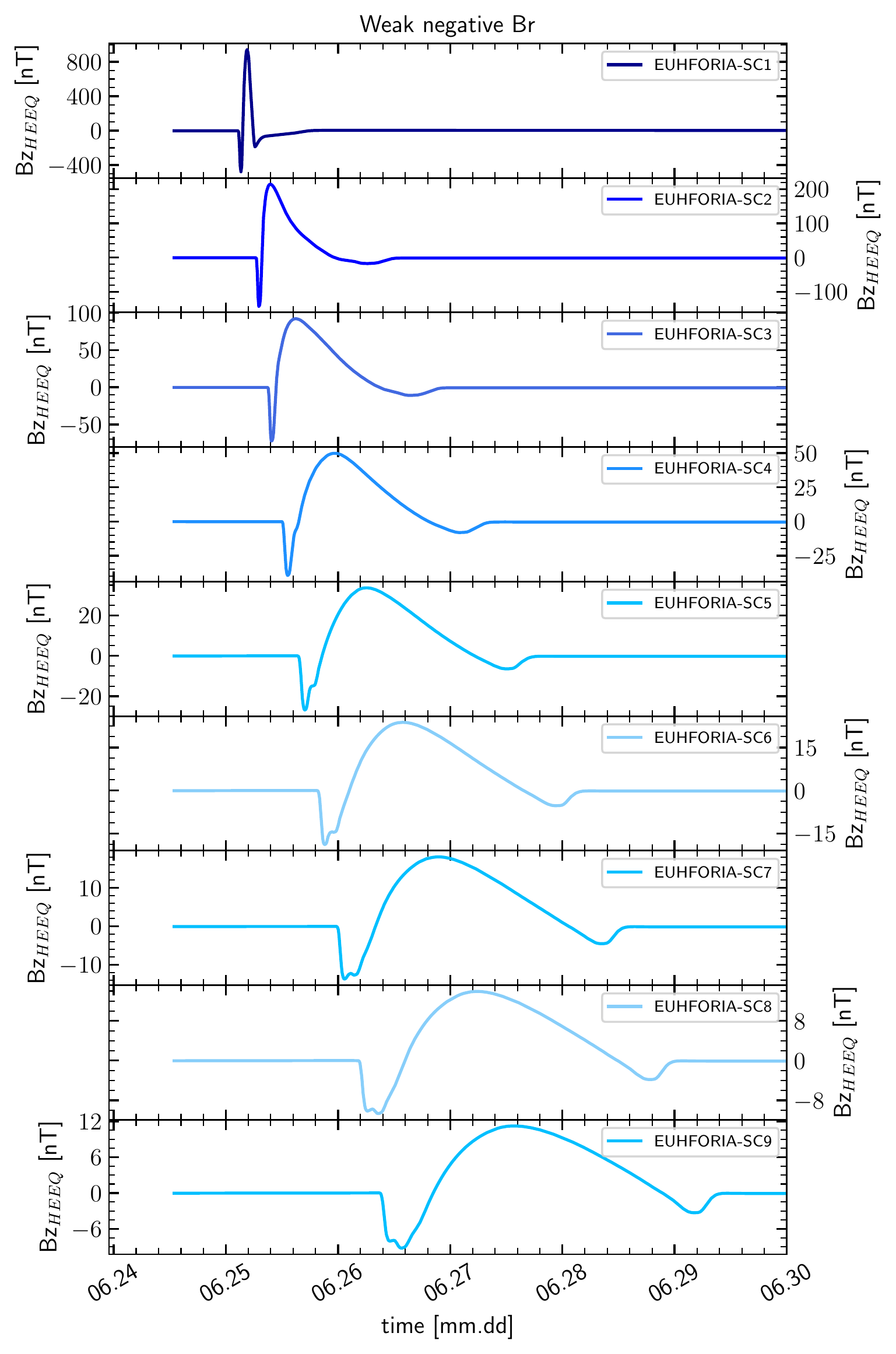}
\includegraphics[height=0.7\linewidth,keepaspectratio]{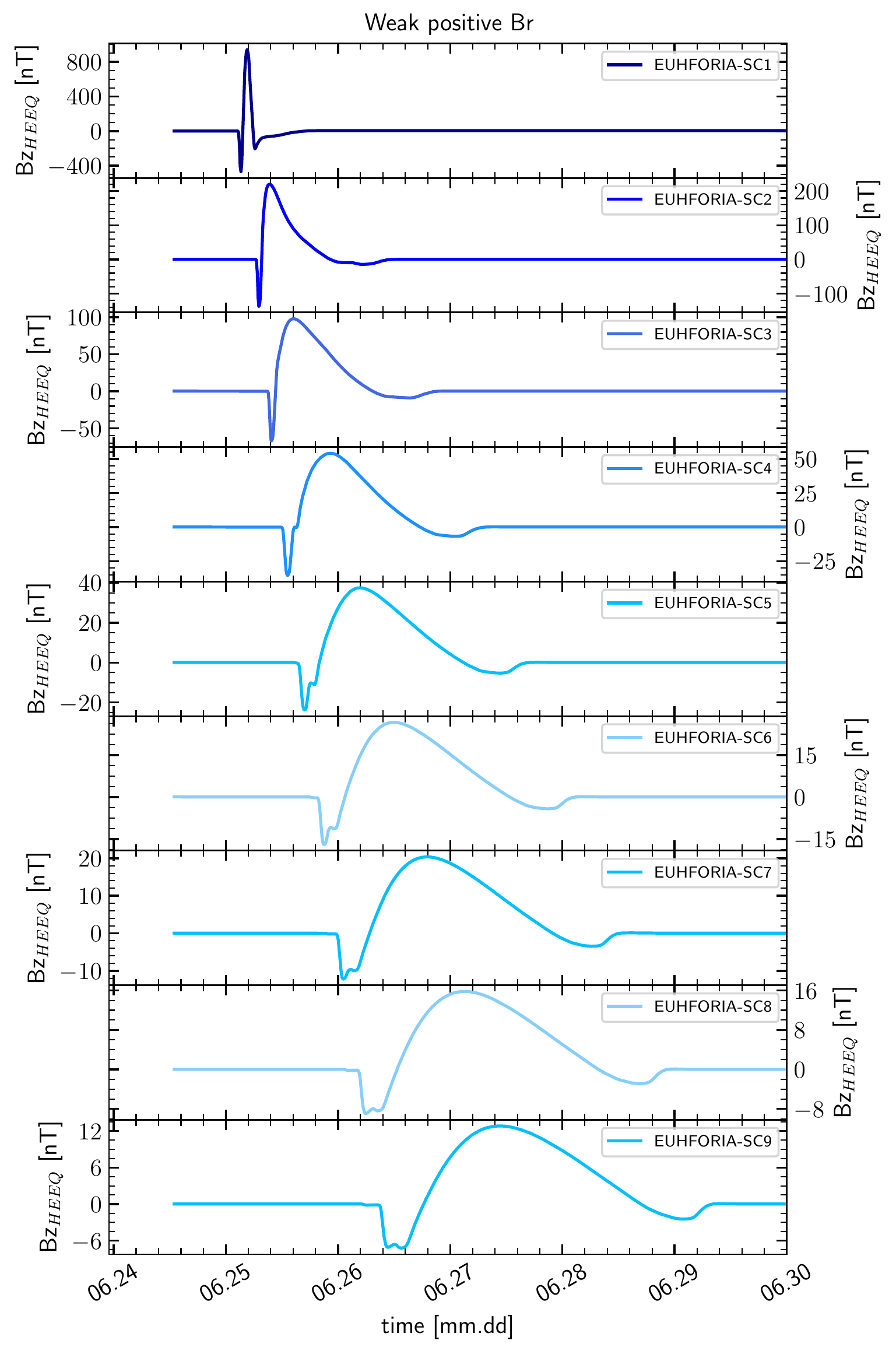}
\caption{Same as Figure \ref{fig:vr_time_series_xweak} but for the $B_z$ component in HEEQ of a spheromak inserted in a weak (left: negative, right:positive) predominantly radial ambient magnetic field.}
\label{fig:vr_time_series_zweak}
\end{figure*}

In our analysis we investigated whether the rotation of the spheromak is manifested in situ at different heliospheric distances. To address this, we placed a set of 9 virtual spacecraft along the Sun-Earth line at heliocentric distances 0.15, 0.3, 0.4, 0.5, 0.6, 0.7, 0.8, 0.9 and 1.0 AU. As the spheromaks in our study were all inserted with initial direction of propagation along the Sun-Earth line, the so-placed spacecraft were supposed to traverse the expanding spheromaks close to their centre, provided that the spheromaks do not get deflected too strongly.

Figures \ref{fig:vr_time_series_xweak} - \ref{fig:vr_time_series_zweak} show the magnetic field components $B_x$, $B_y$, and $B_z$ in the Heliocentric Earth Equatorial (HEEQ) coordinate system as functions of time at the location of the virtual spacecraft, while a CME with initial velocity $v_i=900.0~\mathrm{km}/\mathrm{s}$ and insertion tilt $\theta_i=0^{\circ}$ propagates through the weak negative $B_r$ (left column in each figure) and weak positive $B_r$ ambient field scenario (right column in each figure). Note that the virtual spacecraft and the HEEQ coordinate system are co-rotating with the Earth around the Sun and coincide with the non-rotating (x,y,z)-coordinates used in the previous sections only at the start of the simulations. Over the whole duration of our simulations, the deviation between the two coordinate systems remains, however, small ($\sim3^{\circ}$) and is negligible on length scales of the size of the spheromak.

For the interpretation of the time series in Figures~\ref{fig:vr_time_series_xweak} - \ref{fig:vr_time_series_zweak}, it is instructive to compare the $B_x$, $B_y$, and $B_z$ signal recorded by the virtual spacecraft SC1 for the weak negative $B_r$ background field scenario (left column), with the B-field visualization in Figure~\ref{fig:spheromak_detection}. The $B_y$ time series in Figure~\ref{fig:vr_time_series_yweak}, recorded by SC1, shows nicely the signal one would expect when the spacecraft passes through the center of the spheromak while its axis of symmetry is still more or less aligned with the z-axis: the $B_y$ signal has a single maximum and a single minimum, corresponding to the instants in time where the spacecraft passes through the centers of the red and blue patches in the top-right panel of Figure~\ref{fig:spheromak_detection}. For the spacecraft at larger heliocentric distances, the signal becomes less symmetric. This is partly due to the fact that the spheromak grows and the magnetic flux density steadily drops while the spacecraft are passing through the spheromak and can be expected to have a lower value when the spacecraft reach the back-part of the spheromak. But as the spheromak tilts and slightly drifts, the spacecraft are also no-longer passing through the toroidal flux where it is strongest. It can also be seen, that as the spheromak structure expands while travelling away from the Sun, the leading shock starts to become visible in the time series recorded by the spacecraft SC5-SC9 at larger heliocentric distances. Similarly, the $B_z$ signal recorded by SC1 shows the expected signature of the corresponding poloidal flux: as the front of the spheromak reaches the spacecraft, the $B_z$ corresponds to the poloidal flux that is pointing in negative z-direction there. At the center of the spheromak, the $B_z$ is positive as it now corresponds to the strong poloidal flux passing through there in positive z-direction. Finally, when the tail of the spheromak reaches the spacecraft, the poloidal flux is again in negative z-direction and $B_z$ therefore negative (compare with the 2D field lines shown in the top-right panel of Figure~\ref{fig:spheromak_detection}). If SC1 is traversing the spheromak close to its center and the spheromak's axis of symmetry is still more or less aligned with the z-axis, the $B_x$ signal recorded by SC1 should be flat compared to the $B_y$ and $B_z$ signals, as along the x-axis, both, poloidal and toroidal field should be perpendicular to the x-direction. However, as the spheromak's axis of symmetry is at the location of SC1 already at an angle with the z-axis, the signal is not completely flat, but a significantly smaller amplitude than the corresponding $B_y$ and $B_z$ signals recorded by SC1. For the spacecraft at larger heliocentric distances, the $B_x$ time-series develops a clear negative (for $B_r<0$-scenario) or positive (for the $B_r>0$-scenario) signal for the same duration where the $B_z$ time-series shows the broad maximum corresponding to the poloidal field at the center of the spheromak. The minimum or maximum in the $B_x$ time series therefore indicates that the poloidal field at the center of the spheromak has acquired a component in negative or positive x-direction, which means that the spheromak has tilted in the corresponding direction.

The rotation of the spheromak is only clearly visible in the time series of $B_x$, $B_y$ and $B_z$ recorded by virtual spacecraft, if the spacecraft traverse the spheromak close to its center. For our simulations with the stronger background field scenarios, the latter condition was not well satisfied, due to the larger drift of the spheromaks.

\section{Discussion and Conclusions} \label{sec:conclusion}

We have addressed the phenomenon of spheromak tilting and drifting in the context of modeling the propagation of spheromak CMEs in the inner heliosphere. Spheromak type flux ropes are frequently used to model magnetised CMEs in MHD simulations of the interplanetary space \citep[see for example][]{gibson_time-dependent_1998, vandas3d_mhd_spheromak, manchester_three-dimensional_2004, manchester_modeling_2004, manchester_flux_2014, manchester_simulation_2014, lugaz_numerical_2005, kataoka_three-dimensional_2009, singh_data-constrained_2018, singh_application_2020, singh_modified_2020, jin_data-constrained_2017, shiota_magnetohydrodynamic_2016, verbeke_evolution_2019, scolini_cmecme_2020, asvestari_multispacecraft_2021}. The global magnetic configuration of CMEs is also still an outstanding question, and it is a possibility that some CMEs are spheromaks, or attain a spheromak-like topology via reconnection-driven processes as the eruption evolves from the low to upper corona \citep{gosling_1990, vandas_1993, vandas_1997, Vandas_1998, Farugia_1995, Feng_2021}.

Although, observations do suggest that some CMEs show rotation during the early phases of the eruption in the low solar corona \citep{yurchyshyn_obs_rotate_2008}, this is much more rare in interplanetary space \citep[][]{Isavnin_evolution_2014}. Such \textit{almost rotation-free} expansion in interplanetary space might often not be accurately reproduced by spheromak CME models. A reason for this could be, that spheromaks do already at insertion not sufficiently well reproduce the shape and magnetic field topology a  CME would have if it could be observed \textit{in situ} at that location. Therefore, the spheromak might fit less well into the local ambient magnetic field configuration. The latter might also not be known sufficiently well, due to observational and model limitations, and therefore not be reproduced accurately in simulations. Such a misfit can cause artificial interactions between the magnetic structure of the spheromak and the background magnetic field, which force the spheromak to evolve differently from the \textit{in situ} observed CME. Any spheromak rotation, induced due to lack of alignment between the spheromak axis of symmetry and the ambient field orientation, will affect the magnetic field profiles extracted from the simulation output at a selected locations and has to be taken into account when comparing these profiles with \textit{in situ} observations and/or when used for making space weather forecasts, as this could for example have huge impact on whether a CME is geo-efficient or not.

In Section~\ref{sec:spheromakandtilting} we discussed the different implications of a tilt on stationary, lab-type spheromaks in a constant, homogeneous background magnetic field, or in the presence of stationary boundary surface currents and for expanding space-type spheromaks in a not too strong but otherwise arbitrary ambient magnetic field. In order to remain stationary, the first lab-type spheromak case needs a background magnetic field that is anti-parallel to the spheromak's magnetic moment and therefore counteracts the outward pointing Lorentz force that the spheromak's magnetic field exerts on its own current distribution. This setup is meta-stable as the magnetic moment tends to tilt and align itself with the background magnetic field. If the latter happens, the external field will no-longer counteract the outward pointing Lorentz force that is acting on the spheromak's current distribution, and the spheromak will disintegrate. This is known as the tilting instability \citep{bellan_fundamentals_2000}. In the case of the second stationary lab-type spheromak, the spheromak gets stabilised via induced surface currents on the boundary of the spheromak, and it is then the magnetic field produced by these surface currents which compensates the outward pointing Lorentz forces in the interior of the spheromak. A tilting instability does not occur as the surface currents are induced by the field of the spheromak itself and rotate with the spheromak. The surface currents, however, experience themselves an outward pointing Lorentz force, which needs to be balanced somehow. In a lab-setup, these surface currents are formed on the inner boundary of the conducting walls of the spheromak reactor and therefore confined. In contrast, spheromaks that are inserted in EUHFORIA are not stationary nor confined, but expand. Therefore, there is no meta-stability involved; the expanding spheromak will simply rotate as to align itself with the background magnetic field.

In Section~\ref{subsec:rotation_and_drift} we suggested that the effect of a background magnetic field $\mathbf{B_{\mathrm{sw}}}$ on a spheromak can qualitatively be described by considering the spheromak as a magnetic current-loop dipole, whose magnetic moment $\mathbf{M}$ is then subject to the torque and drift force given by Equations~\eqref{eq:sphtorque} and \eqref{eq:sphdriftforce}.\\
We then monitored the time-evolution of spheromaks with different initial velocities (Section~\ref{subsubsec:diff_initial_speeds}), insertion tilts, and helicities (Section~\ref{subsubsec:diff_insertion_tilts}) in EUHFORIA MHD-simulations using idealised background scenarios with magnetic fields that were set to be predominantly radial at the inner boundary of the modelling domain, either inward or outward pointing. The spheromaks qualitatively behaved as expect from the above-described analogy between a spheromak and a current loop dipole: regardless of the initial tilt of the spheromak, it rotated to align its magnetic moment with the background magnetic field. Also, the spheromaks seemed to have experienced acceleration that deflected their trajectory. Both the rates of rotation as well as deflection angles of the spheromaks were found to be proportional to the strength and sign of the background magnetic field.

We also found that spheromaks start tilting already during insertion.
Namely, the spheromak starts to tilt and drift already before it is completely injected. This indicates that even implementations of the spheromak that remain anchored to the inner boundary experience this tilting which will impede with the model reconstructions at further heliodistances. Let us assume, for example, that a spheromak is inserted in the MHD model with a zero initial tilt (orientation angle), remaining anchored to the inner boundary of the modelling domain, and that it does not experience tilting and drifting. In this case one expects that the structure will only expand in space and that two radially aligned spacecraft will cross through the same section of the spheromak. However, if such an anchored spheromak experiences tilting and drifting it subsequently changes over time its orientation and position in space. This implies that two radially aligned spacecraft will cross each a different section of the spheromak. When inserting spheromaks in MHD models, anchored or not to the inner boundary, one needs to expect that the orientation of the inserted structure will not remain the same.

Finally, in this study we focused on investigating the tilting of spheromaks that had insertion velocities equal to or higher than $v_i=900.0~\mathrm{km}/\mathrm{s}$. However, slower spheromaks were initially also investigated. Spheromaks which were inserted into the modeling domain with an initial velocity that was the same as the velocity of the background solar wind, namely $v_i=450.0~\mathrm{km}/\mathrm{s}$, were subject to significant deformation right after they had detached from the inner boundary. The main reason for the observed dynamics is the incompatibility of the slow insertion speed with the forces arising due to the pressure imbalance between the solar wind and the spheromak. Therefore, modelling slow spheromak needs to be done with caution and by ensuring that the spheromak input parameters are selected carefully so that these effects are reduced. Future investigation on how to reduce these effects by performing a detailed pressure balance assessment is necessary.

The results obtained in this work suggest that spheromaks when inserted in the modelling domain of a heliospheric MHD simulation undergo tilting whenever its magnetic moment is not aligned with the background magnetic field. White light observations strongly indicate that CMEs carry flux-ropes \citep{webb_Howard_cme_review_2013, vourlidas_cme_flux_rope_2013, vourlidas_cme_STEREO_2017} and should therefore also carry a magnetic moment that interacts with the background magnetic field. This implies that CMEs would behave similar to the spheromak and thus undergo tilting. White-light observations of CMEs rarely suggest that CME change orientation during their evolution in the solar corona. However, this does not exclude tilting of the magnetic flux-rope, which is not per se captured by white-light observations. This rises the question: Are observed \textit{in situ} CMEs better aligned with their background magnetic fields than what is the case for the spheromaks in corresponding heliospheric MHD simulations, or do our observations not directly capture such possible flux-rope tilting? At the moment, the answer to the above question remains open, and thus, one can conclude that it is crucial to take the modelled spheromak tilting into account when using spheromaks in space weather and heliospheric modelling of CMEs. 
A more careful embedding of the spheromaks to the ambient solar wind and background magnetic field conditions at the insertion point of the spheromak could improve the agreement between model output and observations. Further research on the spheromak insertion is therefore necessary. Last but not least detailed understanding of the dynamics of magnetized CMEs is required in order to improve space weather predictions of the z component of the magnetic field. To achieve this further observational and theoretical studies of CME eruptions at the Sun and their propagation through the interplanetary medium are necessary.

The methodology used in this manuscript together with a testing of the different weighting factors implemented and their importance in identifying the magnetic centre of mass and axis of symmetry of the spheromak will be discussed in an upcoming paper. Furthermore, in a forthcoming work we will focus on how more complex ambient magnetic field environments, namely the heliospheic current sheet and high speed streams, impact the total tilting angle of the spheromak. In their simulations, \citet{Liu_2019} investigated the interaction of a CME with a co-rotating interaction region, and found that this interaction resulted in drift of the CME. It will be interesting to see whether our upcoming analysis produces similar results. In parallel to investigating the aforementioned complex backgrounds we anticipate to demonstrate the impact of the rotation angle when using the spheromak to model in situ observed CMEs. Last but not least, the effect on the total rotation angle by the modelling set up, namely the resolution, and by the other parameters required as input to the spheromak model, namely the density, temperature, and magnetic flux, are also currently under investigation.

\begin{acknowledgments}

This work was completed under the Project TRAMSEP (Transport Mechanisms of Strong Solar Energetic Particles in Complex Background Solar Wind Conditions), personal funding of E. Asvestari (Academy of Finland Grant 322455). EUHFORIA is developed as a joint effort between the University of Helsinki and KU Leuven. J. Pomoell and E. Kilpua acknowledge funding from the European Union Horizon 2020 research and innovation program under grant agreement 870405 (EUHFORIA 2.0). The results presented here have been achieved under the framework of the Finnish Centre of Excellence in Research of Sustainable Space (Academy of Finland Grant 312390), which we gratefully acknowledge. EK acknowledges the ERC under the European Union's Horizon 2020 Research and Innovation Programme Project 724391 (SolMAG).

\end{acknowledgments}

\begin{appendix}
\section{Locating the spheromak and its geometric axis}
\label{apsec:spheromak_monitoring}
\begin{figure}[htb]
\centering
\includegraphics[width=0.95\linewidth,keepaspectratio]{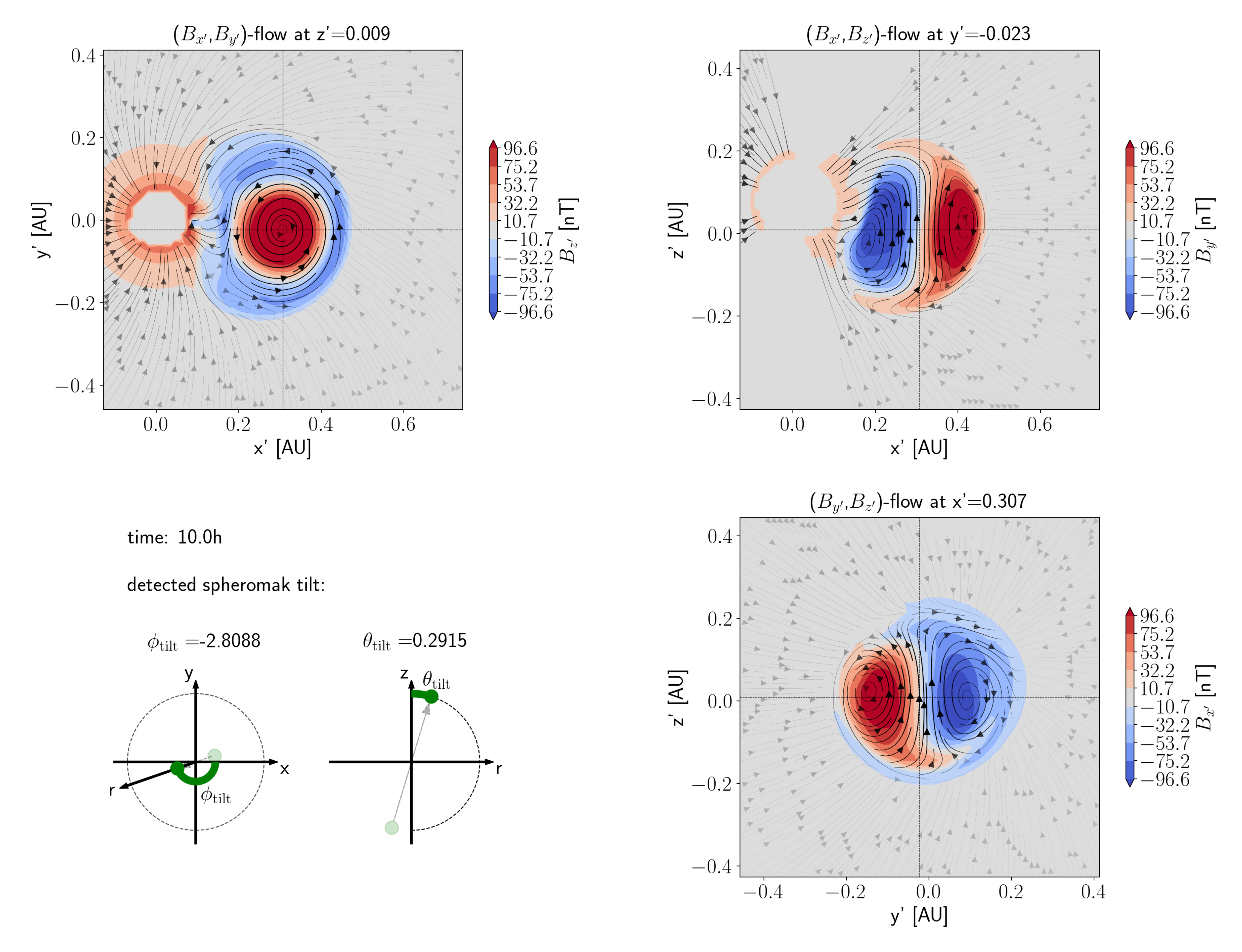}
\caption{Example of an image that is produced by the program that carries out the automated spheromak detection in configuration snapshot of EUHFORIA MHD simulations. The panels show different cuts of the magnetic field $\mathbf{B}$ of a spheromak, 10 hours after having been inserted with initial velocity $v_i=900.0~\mathrm{km}/\mathrm{s}$ into the $B_r<0$, weak background scenario.  The density plot in each panel represents the flow of the vector field $\mathbf{B}$ through the given cut-plane, while the streamlines represent the flow of the $\mathbf{B}$-field component that is parallel to the given cut-plan, with the thickness/opacity of the streamlines representing its magnitude/strength. The detected location of the magnetic centre of mass of the spheromak can be read off from the captions of the individual panels or the location of the "cross-hairs". Primed and unprimed coordinates agree at the centre of rotation. The detected orientation of the spheromak with respect to the original coordinates $\{x,y,z\}$ is summarised in the bottom-left panel where $\theta_{\text{tilt}}$ and $\phi_{\text{tilt}}$ correspond, respectively, to the angles $\theta$ and $\phi$ used in the description of the coordinate transformation from $\{x,y,z\}$ to $\{x',y',z'\}$ in Appendix~\ref{apsec:coordinate_transform}.}
\label{fig:spheromak_detection}
\end{figure}

A spheromak in the modelling domain of an MHD simulation of the inner heliosphere typically has the following two properties:
\begin{enumerate}
\item[a)] its magnetic field carries significantly more energy than the ambient magnetic field at comparable heliocentric distances;
\item[b)] it satisfies, at least locally, the force free condition, $\curl(\mathbf{B})=\lambda \mathbf{B}$, which states that $\curl(\mathbf{B})$ needs to be parallel or anti-parallel to $\mathbf{B}$ (depending on the spheromak's helicity sign). 
\end{enumerate}
Based on this, we decided to locate the volume occupied by the spheromak using the following criteria:
\begin{enumerate}
\item[a)] $|\mathbf{B}(\mathbf{r})|^2$ is bigger than $n_B$ times the median of the values of $|\mathbf{B}(\mathbf{r}')|^2$ at all points $\mathbf{r}'$ that have the same heliocentric distance as $\mathbf{r}$.
\item[b)] $h\,\curl(\mathbf{B})\cdot \mathbf{B}/|\mathbf{B}|>0.9\,|\curl(\mathbf{B})|$, with $h=\pm1$ being the helicity sign. This condition implies that $\curl(\mathbf{B})\neq 0$ and the angle between $\mathbf{B}$ and $h\,\curl(\mathbf{B})$ needs to be smaller than $\arccos(0.9)$.
\end{enumerate}
These determine the volume of the spheromak already pretty well. However, localised disturbed regions in front or behind the spheromak might accidentally also satisfy these criteria. As such regions are typically of much smaller size than the spheromak itself, one can filter them out to some extent by requiring that:
\begin{enumerate}
\item[c)] the magnitude of the magnetic field line curvature, $\text{\boldmath$\kappa$}=(\mathbf{\hat{B}}\cdot\mathbf{\nabla})\mathbf{\hat{B}}$, where $\mathbf{\hat{B}}=\mathbf{B}/|\mathbf{B}|$, must not be bigger than a value $n_R$ times the approximate inverse linear size of the spheromak (as determined with the aforementioned two criteria).
\end{enumerate}
Appropriate values for $n_B$ and $n_R$, depend in general on the simulation input parameters. What is a good value for the parameter $n_B$ depends for example on the relative strength of the background magnetic field and the spheromak's own field. Similarly, the ideal value for $n_R$ will depend on all input parameters that affect the size of localised disturbances that are produced when the spheromak moves through the background solar wind. Good values for these parameters can in principle be determined using statistical methods, but as the focus of this paper is not on the detection of spheromaks but on their dynamics, we found that for the simulation setups considered in this work, the heuristically determined values $n_B=3.0$ and $n_R=3.5$ worked very well.\\
Once the volume $V$ of the spheromak has been mapped out, we determine its magnetic centre of mass:
\begin{equation}
\mathbf{r}_{\mathrm{cm}}=\frac{\int_{V}\mathbf{r}\,w(\mathbf{r})\,\mathrm{d}V}{\int_{V}w(\mathbf{r})\,\mathrm{d}V}\ ,
\end{equation}
where we use the energy content of the part of the $\mathbf{B}$-field that is parallel to its own curl as weight, i.e.:
\begin{equation}
w(\mathbf{r})=\frac{(\curl(\mathbf{B})\cdot \mathbf{B})^2}{2\,|\curl(\mathbf{B})|^2} .\label{eq:densityweight}
\end{equation}
One could also consider switching the roles of $\curl(B)$ and $B$ in Equation \eqref{eq:densityweight}, in order to remove energy contributions from the part of the background magnetic field that might also be parallel to $\curl(B)$. However, these energy contributions are typically rather small, compared to the energy in the field produced by the spheromak. Also, using the square of $\curl(B)$ projected on $B$ as weight, would give unwanted focus on current sheets that might form at the boundary between the spheromak and the background solar wind.\\
Finally, we determine the orientation of the spheromak's geometric axis by assuming that this axis is parallel to the spheromak's magnetic moment, $M$, which is determined as follows:
\begin{equation}
\mathbf{M}=\frac{1}{2}\int_{V} (\mathbf{r}-\mathbf{r}_{\mathrm{cm}})\times \mathbf{J}(\mathbf{r})\,\mathrm{d}V\ ,
\end{equation}
with $\mathbf{J}=\curl(\mathbf{B})/\mu_{0}$.\\
The method utilizes that the system behaves non-relativistically, so that retarded time effects can for example be neglected.\\

In order to assess the quality of the automated spheromak detection with the method described above, the analysis routine produces for each detected case an image as the one shown in Figure~\ref{fig:spheromak_detection}. 
These images show the magnetic field in a coordinate system $\{x',y',z'\}$ that has been rotated around the detected magnetic centre of mass of the spheromak as to align the z'-axis with the detected magnetic moment of the spheromak (see Appendix~\ref{apsec:coordinate_transform} for more details). More precisely, the figures show the $\mathbf{B}$-field along x'y'--, x'z'-- and y'z'--cut-planes that pass through the detected centre of mass of the spheromak. If the determination of the spheromak's centre of mass and magnetic moment have been successful, the panel that shows the x'y'--cut-plane should then always show a top-view of a doughnut-like structure with the 'hole' being in the centre, while the panels for the x'z'-- and y'z'--cut-planes should show corresponding side-views of cuts through the doughnut's centre, with the axis of symmetry being vertical and located in the middle of displayed x' or y' intervals. 
The sequence of these images, produced during the detection, can then be inspected with a viewer or turned into animations that show the time-evolution of the different spheromak types in different background scenarios. This can be done not just for the $\mathbf{B}$-field but for any other quantity such as current density, magnetic field-line curvature, plasma density, pressure, and temperature, etc.

\section{Coordinate transformation}\label{apsec:coordinate_transform}
The dashed coordinates, $(x',y',z')$, used in the illustrations of the properties of the spheromak, correspond to a basis, $\{\mathbf{\hat{x}'},\mathbf{\hat{y}'},\mathbf{\hat{z}'}\}$, that is rotated with respect to the original basis, $\{\mathbf{\hat{x}},\mathbf{\hat{y}},\mathbf{\hat{z}}\}$, in such a way, that $\mathbf{\hat{z}'}$ becomes parallel to the geometric axis of the spheromak and points in the same direction as the magnetic B-field there. This rotation is done in a 'minimal' way, meaning that unnecessary rotations around the z-axis are avoided:
\begin{equation}
(\mathbf{\hat{x}'},\mathbf{\hat{y}'},\mathbf{\hat{z}'})=(\mathbf{\hat{x}},\mathbf{\hat{y}},\mathbf{\hat{z}})\,R(\theta,\phi)\ ,
\end{equation}
where:
\begin{equation}
R(\theta,\phi)=R_{\mathrm{Euler}}(\phi,\theta,-\phi)=\begin{pmatrix}\cos (\theta ) \cos ^2(\phi )+\sin ^2(\phi ) & (\cos (\theta )-1) \sin (\phi ) \cos (\phi ) & \sin (\theta ) \cos (\phi )\\ (\cos (\theta )-1) \sin (\phi ) \cos (\phi ) & \cos (\theta ) \sin ^2(\phi )+\cos ^2(\phi ) & \sin (\theta ) \sin (\phi )\\ -\sin (\theta ) \cos (\phi ) & -\sin (\theta ) \sin (\phi ) & \cos (\theta )\\ \end{pmatrix}\ ,\label{eq:coordtransformmat}
\end{equation}
with $R_{\mathrm{Euler}}(\alpha,\beta,\gamma)$ being the Euler angle rotation matrix in the z-y-z convention, meaning that the first angle refers to a rotation around the z-axis, the second angle to a rotation around the so-obtained new y-axis, and the last angle describes rotation around the new z-axis obtained from the previous two rotations.

The primed coordinates $(x',y',z')$ are then obtained by applying the inverse rotation to $(x,y,z)$, using the spheromak's magnetic centre of mass, $\mathbf{r}_{\mathrm{cm}}=(x_{\mathrm{cm}},y_{\mathrm{cm}},z_{\mathrm{cm}})$, as origin: 
\begin{equation}
\begin{pmatrix}x'\\ y'\\ z'\\\end{pmatrix}=\begin{pmatrix}x_{\mathrm{cm}}\\ y_{\mathrm{cm}}\\ z_{\mathrm{cm}}\\\end{pmatrix}+R^{-1}(\theta,\phi)\,\begin{pmatrix}x-x_{\mathrm{cm}}\\ y-y_{\mathrm{cm}}\\ z-z_{\mathrm{cm}}\\\end{pmatrix}\ .
\end{equation}

\begin{figure}[htb]
\begin{minipage}[t]{0.49\linewidth}
\centering
\includegraphics[height=0.66\linewidth,keepaspectratio]{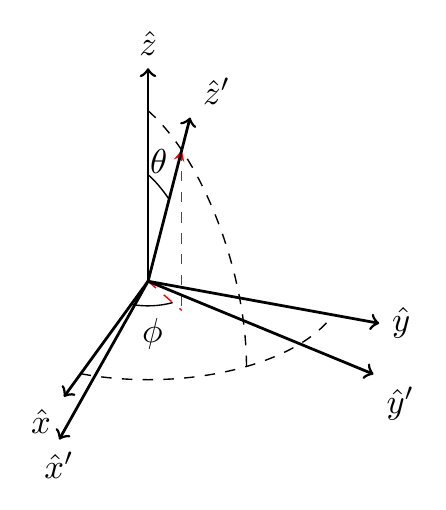}
\caption{Illustration of the action of the basis transformation matrix $R(\theta,\phi)$ from equation \eqref{eq:coordtransformmat}, mapping the original basis $(\mathbf{\hat{x}},\mathbf{\hat{y}},\mathbf{\hat{z}})$ to the primed one: $(\mathbf{\hat{x}'},\mathbf{\hat{y}'},\mathbf{\hat{z}'})$.}
\label{fig:coord_transform}
\end{minipage}
\end{figure}

\section{Magnetic torque and drift in terms of magnetic moment}\label{apsec:torque_and_force_accuracy}
In this section, we discuss in some more detail torque, $\mathbf{\tau}$, and net drift force, $\mathbf{F_{\mathrm{drift}}}$, which a LFF spheromak experiences when subject to a magnetic background field $\mathbf{B_{\mathrm{vac}}}$. In Section~\ref{sec:spheromakandtilting}, we introduced Equations~\eqref{eq:sphtorque} and \eqref{eq:sphdriftforce}, which express these quantities in terms of the magnetic moment, $\mathbf{M}$, of the spheromak. These formulas are, however, only accurate if the spheromak, or more precisely, its current density, $J=\curl(\mathbf{B})/\mu_{0}$ of the spheromak, is restricted to a region that is small compared to the typical length-scale over which the background field $\mathbf{B_{\mathrm{vac}}}$ undergoes significant changes. To make this statement more quantitative, let us first write down the full expressions for $\mathbf{\tau}$ and $\mathbf{F_{\mathrm{drift}}}$, by integrating the local Lorentz force and corresponding torque with which $\mathbf{B_{\mathrm{vac}}}$ acts on the current density distribution $\mathbf{J}(\mathbf{r})$:
\begin{subequations}\label{eq:fulldriftandtorque}
\begin{align}
\mathbf{F_{\mathrm{drift}}}&=\int_{V}\,\mathbf{J}(\mathbf{r})\times \mathbf{B_{\mathrm{vac}}}(\mathbf{r})\,\mathrm{d}V\ ,\label{eq:fulldrift}\\
\text{\boldmath$\tau$}&=\int_{V}\,(\mathbf{r}-\mathbf{r_{\mathrm{cm}}})\times(\mathbf{J}(\mathbf{r})\times \mathbf{B_{\mathrm{vac}}}(\mathbf{r}))\,\mathrm{d}V ,\label{eq:fulltorque}
\end{align}
\end{subequations}
where $\mathbf{r_{\mathrm{cm}}}$ is the "centre of mass" of the magnetic field produced by $J$ and $V$ is the volume occupied by the spheromak. Note that $\mathbf{B_{\mathrm{vac}}}$ is assumed to satisfy $\curl(\mathbf{B_{\mathrm{vac}}})=0$ within $V$. Furthermore, we assume again that the system behaves non-relativistically and retarded time effects can therefore be neglected.

We now Taylor-expand $\mathbf{B_{\mathrm{vac}}}$ in the point $\mathbf{r_{\mathrm{cm}}}$:
\begin{equation}
\mathbf{B_{\mathrm{vac}}}(\mathbf{r})=\mathbf{B_{\mathrm{vac}}}(\mathbf{r_{\mathrm{cm}}})
\quad+\quad((\mathbf{r}-\mathbf{r_{\mathrm{cm}}})\cdot\mathbf{\nabla_{\mathrm{cm}}})\mathbf{B_{\mathrm{vac}}}(\mathbf{r_{\mathrm{cm}}})
\quad+\quad\ldots\ ,\label{eq:bvactaylor}
\end{equation}
where the notation $\mathbf{\nabla_{\mathrm{cm}}}$ means that the derivatives are taken with respect to the components of $\mathbf{r_{\mathrm{cm}}}$ instead of $\mathbf{r}$. Plugging \eqref{eq:bvactaylor} into \eqref{eq:fulldriftandtorque}, we find for \eqref{eq:fulldrift}:
\begin{subequations}\label{eq:expandeddrifandtorque}
\begin{equation}
\mathbf{F_{\mathrm{drift}}}=\int_{V}\mathbf{J}(\mathbf{r})\times \mathbf{B_{\mathrm{vac}}}(\mathbf{r_{\mathrm{cm}}})\mathrm{d}V + \underbrace{\int_{V}\mathbf{J}(\mathbf{r})\times (((\mathbf{r}-\mathbf{r_{\mathrm{cm}}})\cdot\mathbf{\nabla_{\mathrm{cm}}})\mathbf{B_{\mathrm{vac}}}(\mathbf{r_{\mathrm{cm}}}))\mathrm{d}V}_{\mathcal{O}\big(\frac{\lambda_{V}}{\lambda_{\mathbf{B_{\mathrm{vac}}}}}\big)} + \underbrace{\vphantom{\int_{V}}\ldots}_{\mathclap{\mathcal{O}\big(\big(\frac{\lambda_{V}}{\lambda_{\mathbf{B_{\mathrm{vac}}}}}\big)^2\big)}}\ ,\label{eq:expandeddrift}
\end{equation}
and for \eqref{eq:fulltorque}:
\begin{equation}
\text{\boldmath$\tau$}=\int_{V}(\mathbf{r}-\mathbf{r_{\mathrm{cm}}})\times(\mathbf{J}(\mathbf{r})\times \mathbf{B_{\mathrm{vac}}}(\mathbf{r_{\mathrm{cm}}}))\mathrm{d}V
+\underbrace{\int_{V}(\mathbf{r}-\mathbf{r_{\mathrm{cm}}})\times (\mathbf{J}(\mathbf{r})\times(((\mathbf{r}-\mathbf{r_{\mathrm{cm}}})\cdot\mathbf{\nabla_{\mathrm{cm}}})\mathbf{B_{\mathrm{vac}}}(\mathbf{r_{\mathrm{cm}}})))\mathrm{d}V}_{\mathcal{O}\big(\frac{\lambda_{V}}{\lambda_{\mathbf{B_{\mathrm{vac}}}}}\big)}+\underbrace{\vphantom{\int_{V}}\ldots}_{\mathclap{\mathcal{O}\big(\big(\frac{\lambda_{V}}{\lambda_{\mathbf{B_{\mathrm{vac}}}}}\big)^2\big)}}\ .\label{eq:expandedtorque}
\end{equation}
\end{subequations}
The $\lambda_{V}$ and $\lambda_{\mathbf{B_{\mathrm{vac}}}}$ in the underbraces refer to the typical length-scales associated to the volume $V$ of the spheromak (for example the average radius of $V$) and the distance over which $\mathbf{B_{\mathrm{vac}}}$ undergoes significant changes. The ratio of these scales could for example be defined as
\begin{equation}
\lambda_{V}/\lambda_{\mathbf{B_{\mathrm{vac}}}}=\max_{\mathbf{r}\in V}\bigg(\frac{\left|((\mathbf{r}-\mathbf{r_{\mathrm{cm}}})\cdot\mathbf{\nabla_{\mathrm{cm}}})\mathbf{B_{\mathrm{vac}}}(\mathbf{r_{\mathrm{cm}}})\right|}{\left|\mathbf{B_{\mathrm{vac}}}(\mathbf{r_{\mathrm{cm}}})\right|}\bigg)\ .\label{eq:scaleratios}
\end{equation}

Next, we would like to show that the leading terms in these expansions are precisely the Equations~\eqref{eq:sphdriftforce} and \eqref{eq:sphtorque} from Section~\ref{sec:spheromakandtilting}. To evaluate the integrals in Equations~\eqref{eq:expandeddrifandtorque}, we resort to \citep[Section 5.6]{Jackson_1999} and use, that if $\mathbf{J}(\mathbf{r})$ is a localised vector field, and $f(\mathbf{r})$, $g(\mathbf{r})$ are "well-behaved" (no singularities in the integration domain) but otherwise arbitrary functions of $\mathbf{r}$, then:
\begin{equation}
\int \big(f(\mathbf{r})\,\mathbf{J}(\mathbf{r})\cdot\mathbf{\nabla} g(\mathbf{r})\,+\,\underbrace{g(\mathbf{r})\, \mathbf{J}(\mathbf{r})\cdot\mathbf{\nabla} f(\mathbf{r})}_{\mathclap{-f(\mathbf{r})\operatorname{div}(g(\mathbf{r})\,\mathbf{J}(\mathbf{r}))\ \text{(after P.I.)}}}\,+\,f(\mathbf{r})\,g(\mathbf{r})\,\operatorname{div}(\mathbf{J}(\mathbf{r}))\big)\mathrm{d}V=0\ .\label{eq:intidentity}
\end{equation}

The first term in Equation~\eqref{eq:expandeddrift} can now be evaluated by noting that Equation~\eqref{eq:intidentity} with $f(\mathbf{r})=1$ and $g(\mathbf{r})=\mathbf{r}$ tells us that
\begin{equation}
\int \mathbf{J}(\mathbf{r})\mathrm{d}V=0\ ,
\end{equation}
if $\operatorname{div}(\mathbf{J}(\mathbf{r}))=0$, and therefore:
\begin{equation}
\int_{V}\mathbf{J}(\mathbf{r})\times \mathbf{B_{\mathrm{vac}}}(\mathbf{r_{\mathrm{cm}}})\mathrm{d}V = \int_{V}\mathbf{J}(\mathbf{r})\mathrm{d}V \times \mathbf{B_{\mathrm{vac}}}(\mathbf{r_{\mathrm{cm}}})=0\ .
\end{equation}
As the first term in Equation~\eqref{eq:expandeddrift} vanishes, we have to evaluate the second one, which is, unfortunately, a bit more lengthy:
using index-notation (repeated indices are implicitly summed), the term can be re-written as:
\begin{equation}
\int_{V}\mathbf{J}(\mathbf{r})\times (((\mathbf{r}-\mathbf{r_{\mathrm{cm}}})\cdot\mathbf{\nabla_{\mathrm{cm}}})\mathbf{B_{\mathrm{vac}}}(\mathbf{r_{\mathrm{cm}}}))\mathrm{d}V =\mathbf{e}_{i} \int_{V} J_{j}(\mathbf{r}) (\mathbf{r}-\mathbf{r_{\mathrm{cm}}})_{l}\mathrm{d}V\, \epsilon_{i j k} \frac{\partial B_{\mathbf{vac},k}(\mathbf{r_{\mathrm{cm}}})}{\partial r_{\mathrm{cm},l}}\ ,\label{eq:expandeddrifttermtwo}
\end{equation}
where $\{\mathbf{e}_i\}_{i=1,2,3}$ is a set of Cartesian basis vectors and $\epsilon_{i j k}$ is the 3-dimensional Levi-Civita symbol, which is anti-symmetric in all indices and $\epsilon_{1 2 3}=1$. The cross product of two vectors $\mathbf{A}$, $\mathbf{B}$ is the given by:
\begin{equation}
(\mathbf{A}\times\mathbf{B})_i=\epsilon_{i j k}\,A_j\,B_k\ .\label{eq:indexcrossp}
\end{equation}
Now, by setting in Equation~\eqref{eq:intidentity} the functions $f(\mathbf{r})=(\mathbf{r}-\mathbf{r_{\mathrm{cm}}})_l$ and $g(\mathbf{r})=(\mathbf{r}-\mathbf{r_{\mathrm{cm}}})_j$, along with $\operatorname{div}(\mathbf{J}(\mathbf{r}))=0$, we find:
\begin{equation}
\int_{V}(J_j(\mathbf{r})(\mathbf{r}-\mathbf{r_{\mathrm{cm}}})_l+J_l(\mathbf{r})(\mathbf{r}-\mathbf{r_{\mathrm{cm}}})_j)\mathrm{d}V=0\ ,\label{eq:symmintegrand}
\end{equation}
which tells us that if we decompose $J_j(\mathbf{r})(\mathbf{r}-\mathbf{r_{\mathrm{cm}}})_l$ in the integral in Equation~\eqref{eq:expandeddrifttermtwo} into even and odd part, then the even part will vanish:
\begin{equation}
\int_{V} J_{j}(\mathbf{r}) (\mathbf{r}-\mathbf{r_{\mathrm{cm}}})_{l}\mathrm{d}V=\underbrace{\frac{1}{2}\int_{V} (J_{j}(\mathbf{r}) (\mathbf{r}-\mathbf{r_{\mathrm{cm}}})_{l}+J_{l}(\mathbf{r}) (\mathbf{r}-\mathbf{r_{\mathrm{cm}}})_{j})\mathrm{d}V}_{=0}+\frac{1}{2}\int_{V} (J_{j}(\mathbf{r}) (\mathbf{r}-\mathbf{r_{\mathrm{cm}}})_{l}-J_{l}(\mathbf{r}) (\mathbf{r}-\mathbf{r_{\mathrm{cm}}})_{j})\mathrm{d}V\ ,
\end{equation}
and \eqref{eq:expandeddrifttermtwo} becomes:
\begin{multline}
\frac{1}{2}\,\mathbf{e}_{i}\,\int_{V}(\mathbf{J}(\mathbf{r})\times(\mathbf{r}-\mathbf{r_{\mathrm{cm}}}))_{n}\,\mathrm{d}V\,\underbrace{\epsilon_{j l n}\,\epsilon_{i j k}}_{\mathclap{-(\delta_{l i}\delta_{n k}-\delta_{l k}\delta_{n i})}}\,\frac{\partial B_{\mathbf{vac},k}(\mathbf{r_{\mathrm{cm}}})}{\partial r_{\mathrm{cm},l}}\\
=-\frac{1}{2}\,\mathbf{e}_{i}\,\int_{V}(\mathbf{J}(\mathbf{r})\times(\mathbf{r}-\mathbf{r_{\mathrm{cm}}}))_{n}\,\mathrm{d}V\,\bigg(\underbrace{\frac{\partial B_{\mathbf{vac},n}(\mathbf{r_{\mathrm{cm}}})}{\partial r_{\mathrm{cm},i}}}_{=\frac{\partial B_{\mathbf{vac},i}(\mathbf{r_{\mathrm{cm}}})}{\partial r_{\mathrm{cm},n}}}-\delta_{n i}\underbrace{\frac{\partial B_{\mathbf{vac},k}(\mathbf{r_{\mathrm{cm}}})}{\partial r_{\mathrm{cm},k}}}_{\mathclap{=\operatorname{div}_{\mathrm{cm}}(\mathbf{B_{\mathrm{vac}}(\mathbf{r_{\mathrm{cm}}}))=0}}}\bigg)\\
=\bigg(\bigg(\underbrace{\frac{1}{2}\int_{V} (\mathbf{r}-\mathbf{r_{\mathrm{cm}}})\times \mathbf{J}(\mathbf{r})\mathrm{d}V}_{\mathbf{M}}\bigg)\cdot\mathbf{\nabla_{\mathrm{cm}}}\bigg) \mathbf{B_{\mathrm{vac}}}(\mathbf{r_{\mathrm{cm}}})\ ,\label{eq:driftexpandedsecondterm}
\end{multline}
where on the first line, we have used again Equation~\eqref{eq:indexcrossp}, along with the identity,
\begin{equation}
\epsilon_{l m i}\epsilon_{i j k}=(\delta_{l j}\delta_{m k}-\delta_{l k}\delta_{m j})\ ,\label{eq:epsilonidentity}
\end{equation}
where $\delta_{a b}$ is the Kronecker delta; and on the second line, we used the fact that magnetic fields are divergence-free, and that the background field $\mathbf{B_{\mathrm{vac}}}$ by definition satisfies $\curl(\mathbf{B_{\mathrm{vac}}})=0$ within $V$.

We therefore find that the leading contribution to the drift force in Equation~\eqref{eq:expandeddrift} is indeed given by Equation~\eqref{eq:sphdriftforce} from Section~\ref{sec:spheromakandtilting}.

Finding the leading contribution to the torque in Equation~\eqref{eq:expandedtorque} does fortunately not require new formulas or identities. Using again index notation with Expression~\eqref{eq:indexcrossp} and Equation~\eqref{eq:epsilonidentity}, the first term of Equation~\eqref{eq:expandedtorque} can be re-written as:
\begin{equation}
\int_{V}(\mathbf{r}-\mathbf{r_{\mathrm{cm}}})\times(\mathbf{J}(\mathbf{r})\times \mathbf{B_{\mathrm{vac}}}(\mathbf{r_{\mathrm{cm}}}))\mathrm{d}V=\int_{V}\underbrace{\mathbf{J}(\mathbf{r})\,((\mathbf{r}-\mathbf{r_{\mathrm{cm}}})\cdot \mathbf{B_{\mathrm{vac}}}(\mathbf{r_{\mathrm{cm}}}))}_{\mathbf{e}_{j}\,J_{j}(\mathbf{r})(\mathbf{r}-\mathbf{r_{\mathrm{cm}}})_{l}\,B_{\mathrm{vac,l}}(\mathbf{r_{\mathrm{cm}}})}\mathrm{d}V-\int_{V}\underbrace{((\mathbf{r}-\mathbf{r_{\mathrm{cm}}})\cdot \mathbf{J}(\mathbf{r}))}_{\mathclap{\qquad J_{l}(\mathbf{r})(\mathbf{r}-\mathbf{r_{\mathrm{cm}}})_{l}\,B_{\mathrm{vac},j}(\mathbf{r_{\mathrm{cm}}})\,\mathbf{e}_{j}}}\mathrm{d}V\, \mathbf{B_{\mathrm{vac}}}(\mathbf{r_{\mathrm{cm}}})\ .\label{eq:torqueexpanded}
\end{equation}
By using Equation~\eqref{eq:symmintegrand}, we see that the second term in Equation~\eqref{eq:torqueexpanded} has to vanish. The first term can be evaluated analogously to Expression~\eqref{eq:driftexpandedsecondterm} :
\begin{multline}
\int_{V}\underbrace{\mathbf{J}(\mathbf{r})\,((\mathbf{r}-\mathbf{r_{\mathrm{cm}}})\cdot \mathbf{B_{\mathrm{vac}}}(\mathbf{r_{\mathrm{cm}}}))}_{\mathbf{e}_{j}\,J_{j}(\mathbf{r})(\mathbf{r}-\mathbf{r_{\mathrm{cm}}})_{l}\,B_{\mathrm{vac,l}}(\mathbf{r_{\mathrm{cm}}})}\mathrm{d}V=\frac{1}{2}\,\mathbf{e}_{j}\,\int_{V}(\mathbf{J}(\mathbf{r})\times(\mathbf{r}-\mathbf{r_{\mathrm{cm}}}))_{n}\,\mathrm{d}V\,\epsilon_{j l n}\,B_{\mathrm{vac},l}(\mathbf{r_{\mathrm{cm}}})\\
=\bigg(\underbrace{\frac{1}{2}\,\int_{V}((\mathbf{r}-\mathbf{r_{\mathrm{cm}}})\times\mathbf{J}(\mathbf{r}))_{n}\,\mathrm{d}V}_{\mathbf{M}}\bigg)\times \mathbf{B_{\mathrm{vac}}}(\mathbf{r_{\mathrm{cm}}})\ .
\end{multline}
As this first term in Equation~\eqref{eq:expandedtorque} is already non-zero, we stop here and do not evaluate the second term of the expansion. As was the case for the drift force above, we find that also for the torque, the leading contribution to Equation~\eqref{eq:expandedtorque} coincides with the expression in Equation~\eqref{eq:sphtorque} that was used in Section~\ref{sec:spheromakandtilting}. 

We have now verified that the expressions we used in Section~\ref{sec:spheromakandtilting} to describe the tilt and drift that a spheromak experiences when exposed to the magnetic field of the ambient solar wind, are to leading order in $\lambda_{V}/\lambda_{\mathbf{B_{\mathrm{vac}}}}$ correct. 
In order for these leading order expressions to approximate the full force and torque adequately, it is, however, necessary that $\lambda_{V}/\lambda_{\mathbf{B_{\mathrm{vac}}}}\ll 1$. It turns out, that this condition is not extremely well satisfied for the situation of an expanding spheromak that travels through the inner heliosphere. For the background field and spheromak setups used in our simulations, a rough estimate of this ratio indicates that typically $\lambda_{V}/\lambda_{\mathbf{B_{\mathrm{vac}}}}\sim0.6-0.8$ for the faster spheromaks, and for the slowest ones even $\lambda_{V}/\lambda_{\mathbf{B_{\mathrm{vac}}}}\sim0.8-1.1$, although the peak-value here is reached only temporarily.

Furthermore, a CME spheromak in our EUHFORIA runs interacts with the ambient solar wind not only via the described magnetic torque and drift force, but is also subject to pressure gradients (magnetic and thermal) and all sorts of other hydrodynamic effects. While there are not too many alternatives to the magnetic torque from Equation~\eqref{eq:fulltorque} to explain the observed tilting of spheromaks and the alignment of their magnetic moments with the ambient magnetic field, the observed drift of the simulated spheromaks is likely not just due to the magnetic drift force from Equation~\eqref{eq:fulldrift}, but also affected by the above mentioned other effects.

We would like to close the discussion in this appendix by stressing that our analysis of the spheromak tilt is not affected by the fact that Equations~\eqref{eq:sphtorque} and \eqref{eq:sphdriftforce} describe the interaction of a spheromak with an ambient magnetic field only approximately if the background field is sufficiently inhomogeneous. For the simplified situation of a LFF spheromak in a homogeneous background field, discussed in Section~\ref{sec:spheromakandtilting}, the formulas are exact and illustrate that a spheromak can indeed experience a magnetic torque and would also experience a drift force if the background field were slightly inhomogeneous. This makes it plausible that also more complicated magnetic background fields will in general exert torque and drift forces on a spheromak; but, their quantitative description is in general a bit more involved. 

For our measurements of the spheromak tilt, we merely exploited the fact that by computing the magnetic moment associated to the current density, $\mathbf{J}=\curl(\mathbf{B})/\mu_0$, of a spheromak, one gets a quantity that describes the magnetic properties of just the spheromak, with the magnetic background field projected out. Furthermore, the magnetic moment is a relatively stable observable for keeping track of the orientation of the magnetic structure of a spheromak, even if the latter undergoes deformations.

\end{appendix}

\end{document}